\documentclass{aa}
\usepackage[varg]{txfonts}
\usepackage{longtable}
\usepackage{lscape}
\usepackage{graphicx}
\usepackage{xcolor}

\newcommand{\kmprs}  {\mbox{\rm km\,s$^{-1}$}}
\newcommand{\feh} {\mbox{\rm [Fe/H]}}

\newcommand{\FeH} {\mbox{${\rm [\frac{Fe}{H}]}$}}
\newcommand{\CFe} {\mbox{${\rm [\frac{C}{Fe}]}$}}
\newcommand{\OFe} {\mbox{${\rm [\frac{O}{Fe}]}$}}
\newcommand{\MgFe} {\mbox{${\rm [\frac{Mg}{Fe}]}$}}
\newcommand{\AlFe} {\mbox{${\rm [\frac{Al}{Fe}]}$}}
\newcommand{\CaFe} {\mbox{${\rm [\frac{Ca}{Fe}]}$}}
\newcommand{\ScFe} {\mbox{${\rm [\frac{Sc}{Fe}]}$}}
\newcommand{\TiFe} {\mbox{${\rm [\frac{Ti}{Fe}]}$}}
\newcommand{\VFe} {\mbox{${\rm [\frac{V}{Fe}]}$}}
\newcommand{\CrFe} {\mbox{${\rm [\frac{Cr}{Fe}]}$}}
\newcommand{\MnFe} {\mbox{${\rm [\frac{Mn}{Fe}]}$}}
\newcommand{\CoFe} {\mbox{${\rm [\frac{Co}{Fe}]}$}}
\newcommand{\NiFe} {\mbox{${\rm [\frac{Ni}{Fe}]}$}}
\newcommand{\ZnFe} {\mbox{${\rm [\frac{Zn}{Fe}]}$}}
\newcommand{\YFe} {\mbox{${\rm [\frac{Y}{Fe}]}$}}
\newcommand{\ZrFe} {\mbox{${\rm [\frac{Zr}{Fe}]}$}}

\newcommand{\xfe} {\mbox{\rm [X/Fe]}}
\newcommand{\cfe} {\mbox{\rm [C/Fe]}}
\newcommand{\nfe} {\mbox{\rm [N/Fe]}}
\newcommand{\ofe} {\mbox{\rm [O/Fe]}}
\newcommand{\nefe} {\mbox{\rm [Ne/Fe]}}

\newcommand{\mgfe} {\mbox{\rm [Mg/Fe]}}

\newcommand{\alfe} {\mbox{\rm [Al/Fe]}}
\newcommand{\sife} {\mbox{\rm [Si/Fe]}}

\newcommand{\cafe} {\mbox{\rm [Ca/Fe]}}
\newcommand{\scfe} {\mbox{\rm [Sc/Fe]}}

\newcommand{\tife} {\mbox{\rm [Ti/Fe]}}
\newcommand{\vfe} {\mbox{\rm [V/Fe]}}

\newcommand{\cofe} {\mbox{\rm [Co/Fe]}}

\newcommand{\nife} {\mbox{\rm [Ni/Fe]}}

\newcommand{\znfe} {\mbox{\rm [Zn/Fe]}}

\newcommand{\yfe} {\mbox{\rm [Y/Fe]}}
\newcommand{\zrfe} {\mbox{\rm [Zr/Fe]}}
\newcommand{\srfe} {\mbox{\rm [Sr/Fe]}}

\newcommand{\alphafe} {\mbox{\rm [$\alpha$/Fe]}}
\newcommand{\teff}  {\mbox{$T_{\rm eff}$}}

\newcommand{\logteff} {\mbox{${\rm log}\,T_{\rm eff}$}}

\newcommand{\logg}  {\mbox{{\rm log}\,$g$}}

\newcommand{\Msun}  {\mbox{$M_{\sun}$}}

\newcommand{\turb}  {\mbox{$\xi_{\rm turb}$}}

\newcommand{\CI} {\ion{C}{i}}
\newcommand{\OI} {\ion{O}{i}}

\newcommand{\MgI} {\ion{Mg}{i}}

\newcommand{\AlI} {\ion{Al}{i}}

\newcommand{\CaI} {\ion{Ca}{i}}
\newcommand{\CaII} {\ion{Ca}{ii}}
\newcommand{\ScII} {\ion{Sc}{ii}}

\newcommand{\TiII} {\ion{Ti}{ii}}

\newcommand{\VII} {\ion{V}{ii}}

\newcommand{\CrII} {\ion{Cr}{ii}}
\newcommand{\MnI} {\ion{Mn}{i}}
\newcommand{\FeI} {\ion{Fe}{i}}
\newcommand{\FeII} {\ion{Fe}{ii}}
\newcommand{\CoI} {\ion{Co}{i}}
\newcommand{\NiI} {\ion{Ni}{i}}

\newcommand{\ZnI} {\ion{Zn}{i}}

\newcommand{\ZrII} {\ion{Zr}{ii}}
\newcommand{\YII} {\ion{Y}{ii}}

\newcommand{\CeII} {\ion{Ce}{ii}}
\newcommand{\Mv} {\mbox{$M_V$}}
\newcommand{\Vrot}   {\mbox{$V_{\rm rot}$}}

\def\ltsima{$\; \buildrel < \over \sim \;$}
\def\simlt{\lower.5ex\hbox{\ltsima}}
\def\gtsima{$\; \buildrel > \over \sim \;$}
\def\simgt{\lower.5ex\hbox{\gtsima}}

\begin{document}

\title{The intrinsic dispersion of elemental abundance ratios in nearby metal-poor
halo stars
\thanks{Based on data products from observations made with ESO Telescopes
at the La Silla Paranal Observatory under programmes 67.D-0106 and 73.D-0024.}}

\titlerunning{The dispersion of element ratios in halo stars}

\author{P.E.~Nissen \inst{1} \and A.M.~Amarsi \inst{2}}

\institute{Department of Physics and Astronomy, Aarhus University, Ny Munkegade 120, DK--8000
Aarhus C, Denmark  \email{pen@phys.au.dk}
\and Theoretical Astrophysics, Department of Physics and Astronomy, Uppsala University, Box 516, SE--751 20 Uppsala, Sweden}

\date{Received 31 January 2026/ Accepted 28 April 2026}

\abstract
{Information on nucleosynthesis, chemical evolution, and accretion of dwarf galaxies 
at early times in the Milky Way may be
obtained from elemental abundance ratios in halo stars, 
because these ratios depend on the initial mass function (IMF)
for core collapse supernovae (CC\,SNe) and the
role of Type Ia\,SNe in making the elements.  } 
{By determining very precise stellar parameters and abundances for
a sample of intermediate-metallicity halo stars, we want to  estimate the
intrinsic dispersion of various abundance ratios.}
{Differential abundances of C, O, Mg, Al, Ca, Sc, Ti, V, Cr, Mn, Fe, Co, Ni, Zn, Y, and Zr were
determined from high signal-to-noise VLT/UVES spectra
for 25 turnoff stars with $-2.4 < \feh < -1.3$. Effective temperatures were
obtained from profiles of the H$\beta$ line and surface gravities via $Gaia$
parallaxes. The analysis of the spectra were based on 1D model atmospheres assuming
local thermodynamic equilibrium (LTE), but 3D non LTE corrections were applied
for several elements.}
{The dispersion in linear fits to the \xfe -\feh\ relations is around a factor of two 
smaller than found in previous studies. After corrections for measurement errors, 
the 1-$\sigma$ intrinsic dispersion of \xfe\
at a given metallicity is 0.09\,dex for Y and Zr, 0.05-0.07\,dex for C, O, and Al,
0.03-0.05\,dex for Mg, Ca, Sc, Ti, V, Mn, and Zn, and $< 0.03$\,dex for Cr, Co, and Ni.
Strong correlations between the residuals in the \xfe -\feh\ fits are found for the $\alpha$-capture
elements (Mg, Al, Ca, Sc, and Ti) and between the residuals for Y and Zr.} 
{Correlations of the residuals in the \xfe -\feh\ fits with effective temperature
can be explained as due to differential atomic diffusion between elements, but its contribution
to the scatter of \xfe\ is of minor importance. Probably, both stochastic effects 
in sampling the IMF of CC\,SNe and differences in the Type\,Ia to CC\,SNe enrichment ratio
between star-forming regions need to be considered in order to explain the 
intrinsic dispersion of \xfe .}


\maketitle

\section{Introduction}
\label{introduction} 
Trends of abundance ratios as a function of metallicity (\feh \footnote{For two elements, X and Y,
with number densities $N_{\rm X}$ and $N_{\rm Y}$,
[X/Y] $\equiv {\rm log}(N_{\rm X}/N_{\rm Y})_{\rm star}\,\, - 
\,\,{\rm log}(N_{\rm X}/N_{\rm Y})_{\sun}$.}) provide important constraints
on Galactic chemical evolution (GCE) models, but the scatter at a given metallicity is also 
of high interest as shown in high-precision studies of the Galactic disk 
\citep{edvardsson93, fuhrmann98, gratton00, adibekyan12}
and the Galactic halo \citep{cayrel04, cohen04, arnone05}. For example, it came as a surprise 
in these studies that very metal-poor halo stars ($-4.0 <  \feh < -2.5$) showed a 
1-$\sigma$ dispersion of only $\sim \! 0.1$\,dex in the abundances of $\alpha$-capture elements relative
to iron (\alphafe ). In comparison, 
GCE models, for which only a few core collapse supernovae (CC\,SNe) contributed 
to the enrichment of interstellar clouds, predicted dispersions
on the order of 0.4\,dex in \alphafe\ \citep{audouze95, argast00, argast02}. This discrepancy
led to new stochastic GCE models \citep[e.g.][]{karlsson05a, karlsson05b, cescutti08}, 
and further studies of the dispersion of abundance ratios in very metal-poor stars,
of which \citet{li22} is the most precise and extensive. Based on high resolution spectra of 
385 stars covering the metallicity range $-4.3 < \feh < -1.7$ they find 
dispersions in \mgfe , \cafe , and \tife\  $\simlt 0.08$\,dex, whereas the 
dispersion in the abundance of neutron capture elements relative to iron 
is about 0.15\,dex.

Turning to the more 
metal-rich part of Galactic halo, that is stars with metallicities
$\feh > -1.3$, the dispersion in \xfe\ for many elements,
including Na, Mg, Al, Ca, Ti, and Ni, is larger than in the 
case of very metal-poor stars. As found in a high-precision study of abundances of 94 dwarf stars by 
\citet[][hereafter NS10, NS11]{nissen10, nissen11} and confirmed by \citet{hawkins15} and \citet{hayes18}
in an analysis of a larger sample of K giants with APOGEE spectra, the 
more metal-rich part of the halo
splits into two populations: high-$\alpha$ stars born in-situ in the Milky Way
and low-$\alpha$ stars
accreted from dwarf galaxies of which the so-called {\it Gaia}-Sausage-Enceladus (GSE) 
galaxy is the most important \citep{helmi20}. According to NS11,
the mean separation of \mgfe\ at $\feh = -0.9$  between
high-$\alpha$ and low-$\alpha$ stars is 0.22\,dex and the 
measured dispersion for the accreted population, $\sigma \mgfe  = 0.055$\,dex,
is significantly higher than that of the in-situ stars, $\sigma \mgfe  = 0.029$\,dex.
This can be explained by variations in the Type Ia SNe contributions to iron among
dwarf galaxies and formation of high-$\alpha$ stars in regions where only CC\,SNe 
contributed to the chemical evolution. Furthermore, there are also very significant
differences in \xfe\ between in-situ and accreted stars 
for other elements, e.g. Na, Ca, Ti, and Ni, with the dispersion for accreted stars 
always being the highest (e.g.~Figs.~4 and 8 of NS11).

The dispersions in NS11 are challenged by the study of 
\citet{belokurov22}, who used \alfe\ abundance ratios from APOGEE 
and kinematics from $Gaia$ data to separate a large sample of red giant stars 
into in-situ and accreted populations.
According to their Fig. 7, the dispersions of \xfe\ for the two populations are
about the same at $\feh = -1.0$ and about a factor of two higher than the estimated measurement
error, $\sigma_{\rm obs} \simeq 0.02$\,dex. There is, however, a sharp
increase in the dispersion of \xfe\ for the
in-situ stars towards lower metallicities, i.e. $\feh = -1.4$,
especially for N, Al, Mg, and Si, whereas the dispersion is approximately constant
for the accreted stars. This is interpreted as due to a period of chaotic
pre-disc evolution of the Galaxy, dubbed $Aurora$, with bursty star formation.

It is also interesting to compare the measured dispersions in NS11 with those determined 
by \citet{griffith23} in a
high-precision abundance study of a sample of 86 subgiant
stars with $-2.0 < \feh < -1.0$. After correcting for measurement errors,
Griffith et al. derived intrinsic dispersions in linear fits 
of \xfe\ as a function of \feh\ and \alphafe . For the $\alpha$-capture elements, Mg, Si, Ca,
and Ti, they find $\sigma_{\rm intrin} \simeq 0.05$\,dex, i.e. about a factor of two
higher than the measured dispersions in NS11 for the high-$\alpha$ population. 
They ascribe this intrinsic scatter to stochastic sampling of the CC\,SNe progenitor mass
distribution and estimate that ejecta of typically $\sim 50$ CC\,SNe are mixed
over gas masses of $\sim 6 \times 10^4$\,\Msun.

For intermediate metallicity ($-2.5 < \feh < -1.5$) halo stars, high-precision
abundance studies aiming at determining intrinsic dispersions are sparse. The most
precise study is that of \citet{reggiani17}, who determined differential abundance
ratios in a sample of 23 nearby stars situated in the turnoff region of the HR-diagram.
The standard deviation in linear fits of \xfe\ versus \feh\ is close to 0.06\,dex
for Mg, Si, Ca, and Ti, which is of the same order of size as the estimated
measurement error leaving very little room for an intrinsic dispersion in contrast
to the results of \citet{griffith23}. 

It should be noted that these studies typically employ one-dimensional (1D) hydrostatic 
model atmospheres and the assumption of local thermodynamic equilibrium (LTE). 
The wider the span in stellar parameters (effective temperature, surface gravity,
and metallicity in particular), the more potential for errors due to the 1D LTE
assumption to affect the measured dispersion 
\citep[e.g.][]{nissen18,amarsi19c}.
The exact span in stellar parameters at which the 3D non-LTE effects
must be accounted for depends on the line and stellar parameters
in question.  For example, the resonant
\MgI\ 4571.1\,\AA\ intercombination line is a severe case:
changes of around 200\,K in
\teff, 0.25\,dex in \logg, or 0.5\,dex in \feh,
can causes differential 3D non-LTE versus 1D LTE abundance
corrections of around 0.05\,dex
relative to HD\,110621, according to
the models used in the present study.
The subordinate \MgI\ 4703.0\,\AA\ permitted line is less sensitive,
needing changes that are at least twice as large
to result in similar differential corrections.
\citet{griffith23} found that 1D non-LTE effects
do not significantly affect the scatter in their sample; 
however, this has yet to be confirmed with
independent and more recent model atoms as well as with a consistent 3D non-LTE approach.

In this paper, we aim at determining very precise abundance ratios for a sample
of 25 dwarf stars with $-2.4 < \feh < -1.3$ having VLT/UVES spectra with
signal-to-noise S/N$\simgt 200$. By determining the effective temperature (\teff )
with a precision of $\pm 30$\,K, the surface gravity (\logg )
with a precision of $\pm 0.03$\,dex, and by considering 3D non-LTE effects in the abundance analysis, 
errors of determining \xfe\ approach a level of $\pm 0.02$\,dex, which makes it possible
to improve estimates of the intrinsic scatter in order to learn more about the nucleosynthesis of
elements in the early Galaxy.

\section{Sample selection and stellar spectra}
\label{spectra}
The stars have been selected from a sample of 40 dwarf and subgiant stars,
for which we have previously determined sulphur and zinc abundances
\citep{nissen04,nissen07} as well as carbon and oxygen abundances 
\citep{akerman04, fabbian09,amarsi19b}.
Only stars in the turnoff region (6000\,K $\simlt \teff \simlt$6400\,K and
$3.8 < \logg < 4.4$) were included to minimize possible effects of errors
in effective temperature and surface gravity as well as parameter-dependent errors
in the 1D LTE spectroscopic models. Furthermore, stars with $\feh < -2.4$
were excluded because the spectral lines applied are too weak for such stars to
be used for precise abundance determinations. Finally, a few stars with $\feh > -1.3$,
were not included, because they belong to the more 
metal-rich part of the halo, where the
splitting in high-$\alpha$ and low-$\alpha$ stars is already well known. These 
selection criteria leave us with a sample of 25 stars.

The reduction of the UVES spectra is described in the papers cited above. The spectra
include a blue part, 3750--5000\,\AA , previously used to determine 
Fe and Zn abundances and a near-IR part,
6700--10500\,\AA , used to derive C, O, and S abundances. The spectral resolution is
$\lambda / \Delta \lambda = 60\,000$, and  the S/N per spectral resolution 
element is 200 - 300. In this paper, we use only the blue part taking advantage
of the fact that this region contains lines suitable for precise determinations
of the abundances of Mg, Al, Ca, Sc, Ti, V, Cr, Mn, Fe, Co, Ni, Zn, Y, and Zr 
(see Table \ref{table:linelist}).
Equivalent widths (EWs) of these lines were measured with the IRAF $splot$ task
using the same nearby line-free spectral windows to set the continuum
for all stars. Most of the lines have
EW$< 80$\,m\AA\ and could be well fitted with a 
Gaussian profile. Exceptions are the \MgI\ 4703.0\,\AA\ and \AlI\ 3961.5\,\AA\ lines
that were fitted with Voigt profiles.

\begin{figure}
\centering
\resizebox{\hsize}{!}{\includegraphics{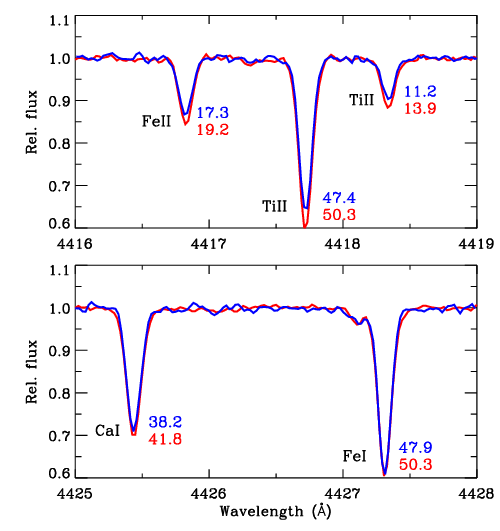}}
\caption{Comparison of Ca, Ti, and Fe lines in spectra of two stars
with similar atmospheric parameters and abundances. The spectrum of
\object{G\,59-27} (\teff \,=\,6232\,K, \logg \,=\,4.22,
\feh\,=\,$-1.90$, \cafe \,=\,0.42, and \tife \,=\, 0.33) is shown with a blue line, and that of
\object{CD\,$-30\,18140$} (\teff \,=\,6241\,K, \logg \,=\,4.15,
\feh \,=\,$-1.84$, \cafe \,=\,0.42, and \tife \,=\, 0.34) with a red line.
For each line the measured equivalent widths for the two stars are given in m\AA\ 
	with an estimated uncertainty of $\pm 0.6$\,m\AA .}
\label{fig:blue.spectra}
\end{figure}

The typical error of the EW measurements for the blue UVES spectra
was estimated by \citet[][see their Fig. 3]{nissen04} to be $\pm 0.6$\,m\AA\ based on a
comparison of EWs measured for six stars that have two spectra observed on different nights.
As an example of the quality of the spectra, Fig. \ref{fig:blue.spectra} shows
a few lines for two stars having similar atmospheric parameters and metallicities.
The small differences between the two stars in the EWs of the lines are mainly due to   
the differences in \logg\ and \feh ; they have almost identical
\cafe\ and \tife\ ratios.

\section{Stellar parameters}
\label{parameters}
\subsection{Effective temperature}
\label{temperature}
The effective temperatures of the stars were adopted from \citet{amarsi19b}, who
determined \teff\ by fitting 3D non-LTE profiles of the H$\beta$ line \citep{amarsi18b}
to the profiles in our spectra. The surface gravities in the present paper are slightly different from
those in \citet{amarsi19b}, because we updated $Gaia$ parallaxes from DR2 to DR3,
and so are the metallicities, but since the H$\beta$ profiles are 
not strongly sensitive 
to \logg\ and \feh\ (with variations in effective temperature
of the order 20\,K due to changes of 0.3\,dex in \logg\ or \feh\ for
metal-poor dwarfs;
see Table 3 of \citealt{amarsi18b}), we did not correct for these small changes. 

\begin{figure}
\centering
\resizebox{\hsize}{!}{\includegraphics{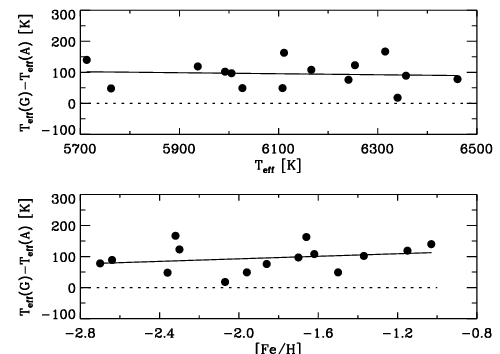}}
\caption{Comparison of effective temperatures determined in \citet{giribaldi21}
from H$\alpha$ profiles and temperatures in \citet{amarsi19b} determined
	from H$\beta$.}
\label{fig:Teff}
\end{figure}

As discussed in \citet{nissen07}, spectra obtained on different nights for a given star
suggest that \teff\ derived from H$\beta$ profiles can be determined 
with an internal 1-$\sigma$ precision of 20\,K. To get an external estimate of the
error, we have compared  the \teff\ values in \citet{amarsi19b} with those determined by 
\citet{giribaldi21} from fitting observed H$\alpha$ profiles to the grid of 3D non-LTE line
profiles in \citet{amarsi18b}. Figure \ref{fig:Teff} shows the comparison for 
15 stars in common. As seen, there is a systematic difference of about 100\,K
demonstrating the well known difficulty in establishing an accurate absolute
\teff\ scale. \citet{amarsi18b} demonstrate that the two features have different dominant
uncertainties:  H$\alpha$ is particularly sensitive to departures from LTE,
which tends to raise the inferred \teff , whereas H$\beta$ shows a large sensitivity 
to the atmospheric structure (as indicated by a higher sensitivity to mixing length 
parameters in 1D models). So, at least from the modelling perspective, 
it is unclear which is to be preferred.

Nevertheless, the standard deviation around the fitted line in Fig.
\ref{fig:Teff} is only 44\,K.
This corresponds to a precision of about 30\,K
for each of the two ways of determining \teff,
assuming that the two \teff\ measurements are independent
and have similar random uncertainties;
this is close to the median
uncertainty of 34\,K in Table 2 of \citet{giribaldi21}.
It is this precision that matters, when discussing differential abundance
ratios for our sample of stars.

\subsection{Surface gravity}
\label{gravity}
The surface gravity of a star was determined from the relation
\begin{eqnarray}
	\log \frac{g}{g_{\odot}}  =  \log \frac{\cal{M}}{\cal{M}_{\odot}} +
	4 \log \frac{\teff}{T_{\rm eff,\odot}} + 0.4 (M_{\rm bol} - M_{{\rm bol},\odot}),
\end{eqnarray}
where $\cal{M}$ is the mass and $M_{\rm bol}$ the absolute
bolometric magnitude. The $Gaia$ DR3 parallax \citep{gaia23} was 
used to derive the absolute visual \Mv\ magnitude from $V$ magnitudes based on Str\"{o}mgren
photometry \citep[][Table B.1]{nissen07}. Interstellar absorption ($A_V$) was estimated from
the colour excess $E(b-y)$ determined from the $(b-y)_0 - \beta$ relation of \citet{schuster89}
using a spectroscopic H$\beta$ index instead of the photometric one in order to
gain in precision \citep{nissen07}. The bolometric correction ($BC$) was derived from
$V - K$ with $K$ magnitudes from \citet{cutri03} using the calibration by 
\citet{casagrande10}, and the
stellar mass was obtained by interpolating in the luminosity - \logteff\
diagram between the Yonsei -Yale evolutionary tracks \citep{yi03}
assuming $\alphafe = 0.4$.

In estimating the uncertainty of \logg , we took into account the parallax error,
$\pm 30$\,K for \teff , and errors of $\pm 0.02$\,mag for $V$ and $A_V$ \citep{nissen07} 
and $BC$ \citep{casagrande10}. These errors propagate to an error of
$\pm 0.05 \cal{M}_{\odot}$ in the stellar mass determined from the evolutionary tracks in the 
luminosity - \logteff\ diagram.
When adding the corresponding errors in \logg\
in quadrature, we obtain errors in the surface gravity around 0.03\,dex.
Thanks to the high precision of the $Gaia$ DR3 parallaxes in combination with
distances from the Sun that are less than 300\,pc for our sample,
the parallax induced error of \logg\ is smaller than the other error sources
contrary to the case in \citet{amarsi19b}, where DR2 parallaxes were used.

\section{Elemental abundances}
\label{abundances}

\subsection{1D LTE analysis}
\label{1DLTE}
Assuming LTE, we used the Uppsala EQWIDTH or BSYN programmes to calculate EWs as a 
function of element abundance for model atmospheres obtained by interpolating
linearly\,\footnote{We have tested that possible errors related to the linear interpolation in the MARCS
grid affect the dispersion of the \xfe -\feh\ relations by less than 0.003\,dex.}
in the standard ($\alphafe = 0.4$) 1D MARCS grid \citep{gustafsson08}
to the \teff , \logg , and \feh\ values of the stars. The abundance corresponding
to a given line can then be determined from the observed EW.
Microturbulence (\turb ) values were obtained by requesting that Fe abundances
derived from $\sim \! 60$ \FeI\ lines have zero slope as a function of
reduced equivalent width ($EW / \lambda$).

The analysis is differential with respect to a standard star, HD\,110621, that has 
a spectrum with a particular high $S/N \sim \! 400$ and a metallicity $\feh = -1.53$. 
We first made an abundance analysis of this star adopting the 
$gf$-values given in Table \ref{table:linelist}. The resulting abundances
are subject to the systematic and statistical errors of the $gf$ values, 
which are typically on the order of $\pm 0.05$\,dex.
For a given element, we then adopted the mean of the derived abundances for HD\,110621
and changed the $gf$-values so that 
all lines of the element provide this mean abundance. 
These revised $gf$-values
were used in the analysis of the other stars, which means that their differential
abundances with respect to HD\,110621  are not affected by errors in the $gf$-values.
It also makes it
possible to determine differential values of \turb\ with a precision of $\pm 0.1$\,\kmprs\
as estimated from the error of the slope of Fe abundance versus reduced EW.

Van der Waals broadening due to collisions with
neutral hydrogen and helium atoms was based on the ABO (Anstee, Barklem, O'Mara)
calculations \citep[see][]{barklem16} for lines marked with "A" in Table  \ref{table:linelist}.
For lines marked with "U", the \citet{unsold55} approximation with an enhancement
factor of 1.5 was applied.

Hyperfine structure ({\em hfs}) broadening was taken into account for the lines of Al, Sc, V, Mn, and Co 
adopting data from the following ources: Al \citep[][VALD\footnote{{\tt https://vald.astro.uu.se}}]
{brown99}, Sc \citep{lawler19},
V \citep{wood14a}, Mn \citep{mcwilliam95},
and Co \citep{lawler15}. Whereas the effect on the derived abundances of Al, Sc, V, and Mn 
is negligible small, due to  a relatively small {\em hfs} splitting,
the effect on the Co abundances derived from the 4121.3\,\AA\ \CoI\ line is important. 
The standard deviation of the linear fit to \cofe\ versus \feh\ decreases from 0.036\,dex 
without {\em hfs} to 0.024\,dex when {\em hfs} splitting is included. Also the slope of \cofe\ vs. \feh\
is significantly affected.

The error in the abundance determination of an element X for a given star
arising from the uncertainties in the EW measurements
was calculated as
$\sigma _{\rm X} = \sigma _{\rm X,line} / \sqrt{N_X}$, where $\sigma _{\rm X,line}$ is
the standard deviation in the line-to-line scatter of the abundance of X
and $N_X$ is the number of lines\,\footnote{For the elements Al, Co, and Zr with only 
one line available, $\sigma_{\rm X,line}$
was  estimated from the  error of the equivalent width corresponding to the S/N of the spectrum.}.
As the S/N of the spectra and the number of lines applied for the various stars are 
much the same, we calculated an average error $\langle \sigma _{\rm X} \rangle$.
This average error was employed
for the whole sample, on a per-element basis.
The corresponding average error of \xfe\ is then 
\begin{eqnarray}
\langle \sigma \xfe \rangle = 
	\sqrt{\langle \sigma _{\rm X} \rangle^2 + \langle \sigma _{\rm Fe} \rangle^2}
\end{eqnarray}
To this we added in quadrature the errors of \xfe\ arising from the uncertainties
in \teff\ ($\pm 30$\,K), \logg\ ($\pm 0.03$\,dex) and \turb\ ($\pm 0.1$\,\kmprs ), 
which are also much the same for all stars, due to their similarity in the
parameters. In this way we arrive at a representative 1-$\sigma$ "observational" error,
$\sigma_{\rm obs}$, of \xfe\ to be used when calculating the intrinsic dispersion
of linear fits of \xfe\ as a function of \feh ; see Sect. \ref{results}.

\subsection{Corrections for 3D (non)-LTE effects}
\label{3DnonLTE}

For some elements, 3D LTE or 3D non-LTE radiative transfer calculations have been used 
to obtain corrections of the derived 1D LTE abundances.
In the following, these calculations are described and the impact of the corrections on
the abundances is discussed. To be short we use the notations:

\smallskip
3N\,--\,1L = $A$(3D non-LTE) -- $A$(1D LTE)

\smallskip
3L\,--\,1L = $A$(3D LTE) -- $A$(1D LTE),

\noindent
where $A$ denotes the logarithmic abundance.
We stress that in the 
\xfe -\feh\ plots that are presented in subsequent sections, 
the abundance of the reference element Fe
appearing in both \xfe\ and \feh\ is based on 3D LTE abundances
inferred from \FeII\ lines, regardless of the type
of model used for the diagnostic element X.
As the 3D corrections for Fe are not a strong function
of stellar parameters and the non-LTE effects
are negligible (Sect.~\ref{Fe}), the trends
would not be much affected if 1D LTE
or 3D non-LTE Fe abundances were instead
adopted in the ratio \xfe.

\smallskip

The 3D LTE and 3D non-LTE abundance corrections are based on the \texttt{Stagger} grid of 
3D stellar atmosphere models by \citet{magic13}; see also \citet{diaz24}
and \citet{stein24}.  For C, O, Mg, Ca, and Fe, these models were post-processed using 
the 3D non-LTE code \texttt{Balder} \citep{amarsi18a}, an offshoot from \texttt{Multi3D} 
\citep{botnen99, leenaarts09}.  For the ionized lines of Ti, corrections were instead 
calculated using the 3D LTE code \texttt{Scate} \citep{hayek11} with updates to the 
input data described in \citet{amarsi21}.  Historically, the accuracy of non-LTE modelling 
has been hampered by unrealistic treatments of the inelastic collisions with neutral hydrogen;
these limitations could potentially introduce additional scatter into the abundances 
\citep[e.g.][]{caliskan25}. In the current study, all of the model atoms used for 
3D non-LTE corrections employ modern descriptions for these processes, 
that are based on asymptotic model approaches \citep[e.g.][]{barklem21} rather than 
the outdated Drawin approximation.

The 3D models naturally take into account line broadening caused by stellar granulation 
without additional free parameters, whereas in 1D the effect of the small scale velocity field
on the equivalent widths is accounted for with the microturbulence `nuisance' parameter 
\citep[e.g.][]{ludwig16}.  The 3N\,--\,1L and 3L\,--\,1L abundance corrections are a function of the 
microturbulence parameter adopted in the 1D LTE analysis.  This ensures that the 3D abundances
are independent of the choice of the 1D microturbulence parameter.

\subsubsection{Carbon and oxygen}
\label{CO}
C abundances were derived from up to seven \CI\ lines in the 9000 - 9500\,\AA\ spectral range
and O abundances from the \OI\ triplet at 7772 - 7775\,\AA\ using EWs measured in 
VLT/UVES spectra by \citet{nissen02}, \citet{akerman04}, and \citet{fabbian09}. 
The same data were used by \citet{amarsi19b} to derive C and O abundances for 
our sample, but we have updated their results to the new parameters of the stars. 

3N\,--\,1L corrections were adopted from \citet{amarsi19c} that are based on model atoms 
described in \citet{amarsi18a} and \citet{amarsi19a}. The provided routines were used to 
interpolate onto the updated stellar parameters (\teff , \logg , \feh, \turb , and $A$(C) or $A$(O)) 
used in the present work.
In the case of carbon, the 3N\,--\,1L corrections range from about $-0.05$\,dex at 
$\feh = -1.3$ to $-0.15$\,dex at $\feh = -2.3$ with a dispersion of 0.025\,dex. The corrections for oxygen
have the opposite trend ranging from about $-0.13$\,dex to $-0.07$\,dex over the same metallicity 
range and a dispersion of only 0.017\,dex for our sample.
All three \OI\ lines arise from the same multiplet,
and the 3D non-LTE effects for the lines of \OI\ therefore
have the same sign,
are of similar magnitude, and have similar sensitivities to 
stellar parameters.
This is also the case for the set of \CI\ lines,
for which six of them are from the same multiplet,
while the seventh has 
excitation energy that is close to that of the other six
(see Table 1 of \citealt{amarsi19a}).
Consequently, the
line-to-line scatter is essentially unchanged after introducing the 3N\,--\,1L corrections
and so are the standard deviations of the fits of \xfe\ vs. \feh .  
The slopes of the \xfe -\feh\ relations are, however, significantly affected due to the 
rather strong dependence of the 3N\,--\,1L corrections on \feh .

\subsubsection{Magnesium}
\label{Mg}
The Mg abundances were derived from two \MgI\ lines, that is the 4571.1\,\AA\ line with an excitation
potential of the lower level $\chi_{\rm exc} = 0.00$\,eV and the 4703.0\,\AA\ line
with $\chi_{\rm exc} = 4.34$\,eV. The first line is quite weak with the EWs ranging from
$\sim \! 3$\,m\AA\ to $\sim \! 30$\,m\AA\ over the metallicity range of our stars, whereas
the EW of the 4703.0\,\AA\ line ranges from $\sim \! 40$\,m\AA\ to $\sim \! 130$\,m\AA .

\begin{figure}
\centering
\resizebox{\hsize}{!}{\includegraphics{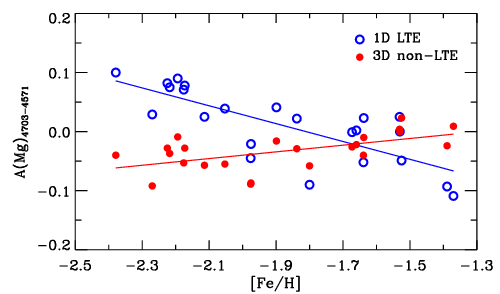}}
\caption{Difference of Mg abundances derived from the 4703.0\,\AA\ and 4571.1\,\AA\ lines as a
	function of \feh\ normalized to zero for the standard star, HD\,110621.}
\label{fig:Mg4703-4571}
\end{figure}

3N\,--\,1L corrections were based on the calculations presented in \citet{matsuno24}, who 
adopt the model atom described in \citet{asplund21}. These data were interpolated onto 
the stellar parameters \teff , \logg , \turb , and $A$(Mg).
The correction for the 4571.1\,\AA\ line changes strongly with \feh , that is
from $\sim \! -0.05$\,dex at $\feh = -1.3$ to $\sim \! +0.15$\,dex at $\feh = -2.3$. 
The correction for the  4703.0\,\AA\ line is more constant at a level
of 0.10 to 0.15\,dex depending on \teff\ and \logg . Interestingly, the agreement
of the Mg abundances derived from the two lines is improved after applying 
the 3N\,--\,1L corrections  as shown in Fig. \ref{fig:Mg4703-4571}. The trend of the difference
in $A$(Mg) is flatter and the dispersion is smaller in 3D non-LTE compared to 1D LTE;                       
$\langle \sigma_{\rm Mg} \rangle$ calculated from the line-to-line scatter
drops from 0.026\,dex in 1D LTE to 0.019\,dex in 3D non-LTE.
Despite of this improvement in $\langle \sigma_{\rm Mg} \rangle$, the standard deviation of the
fit to \mgfe\ versus \feh\ does not change significantly due to the 3N\,--\,1L corrections. 
However, this does not mean that the 3N\,--\,1L corrections are unimportant. As discussed 
in Sect. \ref{results}, the correlation of the residuals in the \mgfe - \feh\ relation with the 
residuals for the other $\alpha$-capture elements is much stronger, when the corrections are included.

\subsubsection{Aluminium}
\label{Al}

The only lines in our spectra available for determining Al abundances are the resonance doublet at 
3944.0, 3961.5\,\AA . The 3944.0\,\AA\ line is distorted by \CrII\ and \CeII\ lines 
\citep{ernandes25} and unsuitable for 
precise abundance determinations. The 3961.5\,\AA\ line has a nice profile, but it falls in the
broad wings of the of the hydrogen H$\epsilon$ line at 3970.1\,\AA\ and the \CaII\ line at
3968.5\,\AA\ that depress the flux at the wavelength of the \AlI\
line by 5 - 10\% for our stars. To take this into account, we measured the EW of the \AlI\ line
relative to the flux in the wings of the H$\epsilon$ and \CaII\ lines and calculated
a corresponding EW from a synthetic spectrum for which the contribution to the line absorption coefficient
from the two broad lines is included. The profile of H$\epsilon$ was calculated as
in \citet{barklem02} and for the calculation of the \CaII\ line we adopted log\,$gf = -0.18$
from the NIST database \citep{nist24} and Ca abundances from the present work.

3N\,--\,1L corrections were computed specifically for this work, employing the model atom of 
\citet{nordlander17}.  The calculations were for 3D models with \teff\ of approximately 6000K and 6500K, 
\logg\ of 4.0 and 4.5, \feh\ of $-1$, $-2$, and $-3$. For the post-processing calculations, 
\alfe\ was varied between $-0.8$ dex and +0.4 dex, in steps of 0.4 dex. The 3D models were 
downsampled horizontally by a factor of three (number of meshpoints $240^2 \rightarrow 80^2$) 
and upsampled in the vertical direction, and post-processing was performed for five snapshots; 
this approach is sufficient for determining reliable abundance corrections based on equivalent widths 
\citep{diaz24}. For the mean radiation field, the calculations employed $26$ rays across 
the unit sphere, based on the $8$-point Lobatto quadrature between $-1\leq\mu\leq1$ and the
$4$-point trapezoidal quadrature between $0\leq\phi\leq2\pi$; whereas for the emergent 
disc-integrated flux, the calculations employed $41$ rays across the unit hemisphere, 
based on the $7$-point Lobatto quadrature between $0\leq\mu\leq1$ and the $8$-point
trapezoidal quadrature between $0\leq\phi\leq2\pi$. 1D LTE and 1D non-LTE calculations were 
also performed on \texttt{ATMO} model atmospheres (the 1D equivalent of the \texttt{Stagger} models); 
see Appendix A of \citet{magic13}. These were in general performed in an analogous way, 
but covering a wider and finer grid of [X/Fe] and also introducing a microturbulence parameter,
\turb = 0, 1, 1.5, and 2 \kmprs.

The EW of the 3961.5\,\AA\ line ranges from about 60\,m\AA\ to more than 150\,m\AA\ over the
metallicity range of our stars. At $\feh = -2.3$, the 3N\,--\,1L correction
is as large as $\sim \! 0.6$\,dex decreasing to $\sim  \! 0.4$\,dex at $\feh = -1.3$,
and at a given metallicity there are variations in 3N\,--\,1L up to 0.1\,dex depending on \teff\
and \logg . The 3N\,--\,1L corrections have little effect on  
the dispersion of the \alfe -\feh\ relation, but as in the case of Mg, the corrections 
are very important for the correlations of the \alfe\ residuals with the residuals in the
\xfe -\feh\ relations for $\alpha$-capture elements (see Sect. \ref{results}).

\citet{skuladottir25} and \citet{ernandes25} have recently used the 3961.5\,\AA\ \AlI\
line to derive Al abundances for the NS10 high-$\alpha$ and low-$\alpha$ 
stars, that span the metallicity range $-1.6 < \feh < -0.7$. Based on the grids of
\citet{ezzeddine18}, they find 1D non-LTE corrections on the order of 0.2\,dex, which agrees
well with the 1N\,--\,1L corrections we derive for the more metal-rich stars in our sample.
The 3D effects add, however, another $\sim \! 0.2$\,dex to the corrections, so that
our 3N\,--\,1L corrections are on the order of 0.4\,dex at $\feh = -1.3$. As a 
consequence of this, we find a level of \alfe\ of about $-0.35$\,dex 
(see Sect. \ref{results}) compared to 
$-0.55$\,dex for the low-$\alpha$ stars in \citet{ernandes25}. 
We note that our level of \alfe\ agrees 
fairly well with the level of $\alfe = -0.3$ for low-$\alpha$ stars obtained
from APOGEE spectra \citep{hayes18} in 1D LTE \citep{garciaperez16}. 
The deeper-forming infrared \AlI\ lines used in APOGEE
\citep{shetrone15} show only mild 1D non-LTE effects
in metal-poor giants
(e.g.~Fig.~13 of \citealt{nordlander17}, and Table A3 of \citealt{lind22}).
However, 3D non-LTE effects remain to be investigated.

\subsubsection{Calcium}
\label{Ca}

The Ca abundances were derived from five weak \CaI\ lines (see Table \ref{table:linelist}).
Like for Al, 3N\,--\,1L corrections were computed specifically for this work, 
employing the model atom described in \citet{asplund21} that was shown to give quite consistent 
results between the neutral and ionized lines at low metallicity by \citet{lagae23}. 
The calculations were performed in almost the same way as those for Al (see Sect. \ref{Al}), 
albeit with a smaller range of abundances (\cafe\ between -0.4 dex to +0.4 dex for the 3D runs),
owing to the larger computational cost and lower severity in the abundance corrections.

The  3N\,--\,1L corrections are much the same for the five \CaI\ lines and
the line-to-line scatter is only slightly improved by the corrections,
that is from $\langle \sigma_{\rm Ca} \rangle$ = 0.013\,dex in 1L
to 0.012\,dex in 3N. The mean 3N\,--\,1L correction increases with decreasing \feh\
but the dependence on \teff\ and  \logg\ is negligible. Consequently, there is no significant effect 
on the dispersion in the \cafe -\feh\ relation or on the correlation of the residuals with 
the residuals for other \xfe -\feh\ relations.

\subsubsection{Titanium}
\label{Ti}

The Ti abundances were derived from up to 16 \TiII\ lines with EWs ranging from 8 to 67\,m\AA\ for
the standard
star HD\,110621. Using 1D model atmospheres, \citet{mallinson22, mallinson23, mallinson24} 
have skown that non-LTE effects on Ti abundances derived from \TiII\ lines are very small.
We assume that this is also the case for 3D models and have calculated 3L\,--\,1L corrections
on a slightly more expanded grid than for Al and Ca, also covering models with \teff =6000\,K 
and \logg =3.5, as well as \teff =5500\,K and \logg = 3.5, 4.0, and 4.5. 
The calculations were performed on a line-by-line basis, varying the oscillator strength in steps 
of 0.2\,dex. The resulting 3L\,--\,1L corrections are small at a level of 0.01\,dex with no
dependence on \teff\ and \logg . Hence there is no significant effect of the corrections on the 
slope and scatter of \tife\ as a function of \feh .

\subsubsection{Iron}
\label{Fe}

Iron abundances were derived from 14 \FeII\ lines with EWs ranging from 9 to 70\,m\AA\ in the standard
star HD\,110621. 
As is well documented \citep[e.g.][]{lind12}, \FeII\ lines
have the advantage of being much less sensitive to departures
from LTE than \FeI\ lines.
As discussed in \citet{amarsi16} and 
\citet{amarsi22}, the 3D non-LTE versus 3D LTE abundance differences
for such lines are
only around 0.02\,dex for F-dwarfs
with \feh$=-2$; the differential corrections relative to the standard star
are even smaller.
In contrast, for \FeI\ lines the corresponding differences
usually exceed a factor of two (0.3\,dex).
We have therefore applied 3L\,--\,1L corrections from \citet{amarsi19c}, using the same 
interpolation tools as for C and O (see Sect. \ref{CO}).
The mean corrections for the 14 \FeII\ lines are nearly constant at a level of 0.10\,dex with 
a dispersion of only 0.006\,dex for our sample of stars.

\subsubsection{Other elements}
\label{Other}

For five of the remaining elements, Sc, V, Cr, Y, and Zr, lines from the ionized
majority species were applied.
As majority species are less sensitive to the effects
of overionisation and overrecombination, and 
these elements are all found in the d-block with dense energy
structures, one expects small non-LTE effects. 
As the 3D LTE corrections were found to be small
for  \TiII\ (also in the d-block), we speculate
that overall there are probably only small
3D non-LTE effects on the trend and scatter of the \xfe -\feh\ relations.
An exception to this argument is \CrII , for which the ground state has the more stable
half-filled 3d$^5$ configuration, and consequently the energy levels have much
larger separations.
For the other d-block elements Mn, Co, Ni, and Zn, only lines of neutral atoms were 
available and we cannot exclude
that 3N\,--\,1L corrections affect the slopes and scatter of the
\xfe -\feh\ relations significantly. Hence, the results for these elements should 
be considered as preliminary.

\section{Results}
\label{results}

\begin{figure*}
\centering
\includegraphics[width=18.0cm]{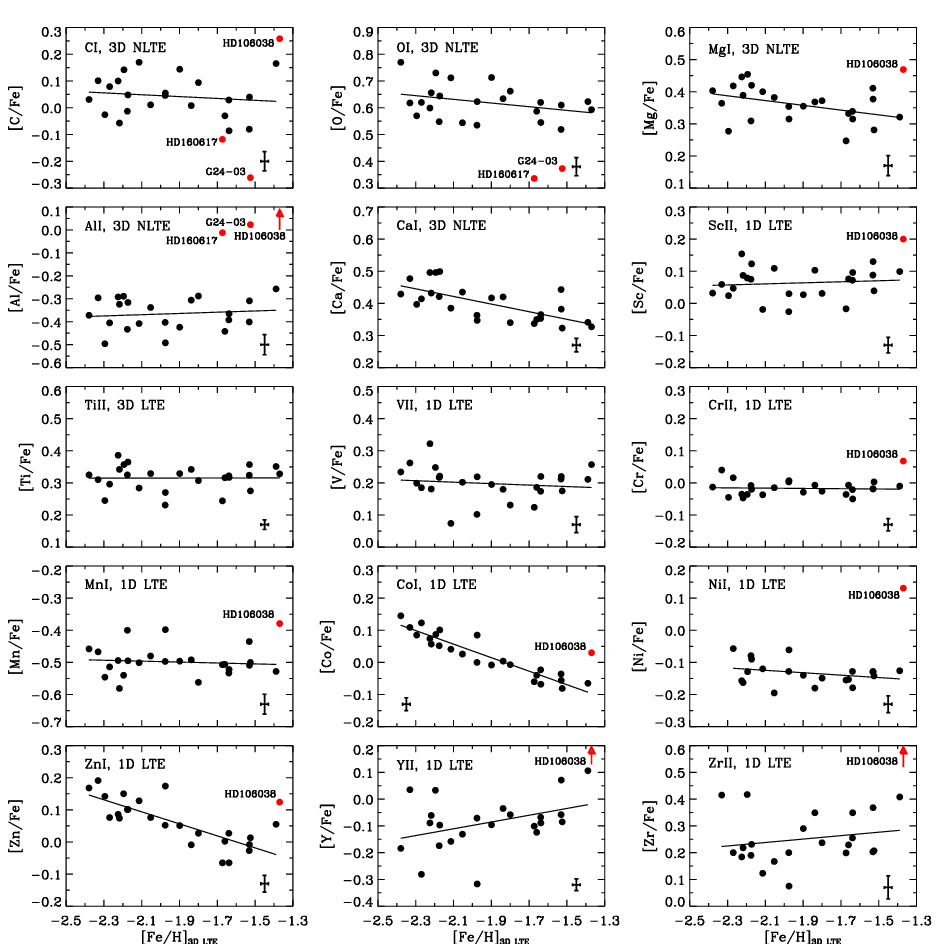}
\caption{\xfe\ as a function of \feh\ for 15 elements. 
On each panel, the atomic species of the spectral lines used for deriving abundances
and the method of analysis are shown. The straight lines show least-squares fits 
to stars marked by black filled circles. Stars marked by red filled circles for some elements 
are considered peculiar and are not included in the fits. In the Al, Y, and Zr panels, HD\,106038
	falls above the upper frame, which is indicated by a red arrow. The error bars 
	shown in this and the following figures refer to the values of $\sigma_{\rm obs}$
	given in Table \ref{table:fit.feh}.}
\label{fig:xfe-feh}
\end{figure*}

The derived atmospheric parameters, \feh\ abundances, and \xfe\ ratios are given
in Table A2 and \xfe\ as a function of \feh\ is shown in Fig. \ref{fig:xfe-feh}.
The lines in this figure show 
least-squares linear fits, $\xfe = a + b \cdot \feh$, to the data after
excluding a few peculiar stars for
some abundance ratios; see discussion of these stars in Sect. \ref{Binaries}. 
The slopes of the fits are given in  
Table \ref{table:fit.feh}. For most elements, the slopes are not significantly different 
from zero; exceptions are Ca, Co, and Zn for which there is a clear decreasing trend of 
\xfe\ with \feh .

Table \ref{table:fit.feh} also lists the standard deviation of the fits, $\sigma_{\rm fit}$,
together with the estimated observational error of \xfe ,
$\sigma_{\rm obs}$ (see Sect. \ref{1DLTE}), and the 
intrinsic dispersion derived as  
\begin{eqnarray}
\sigma_{\rm intrin} = \sqrt{\sigma_{\rm fit}^2 - \sigma_{\rm exp}^2}\,,
\end{eqnarray}
where $\sigma_{\rm exp}$ is the scatter around the fitted line expected from the
observational errors 
\begin{eqnarray}
	\sigma_{\rm exp}^2 = \sigma_{\rm obs,[X/Fe]}^2 + 
				 b^2 \cdot \sigma_{\rm obs,[Fe/H]}^2, 
\end{eqnarray}
where $\sigma_{\rm obs,[Fe/H]} = 0.015$\,dex.

\begin{table}
\centering
	\caption[ ]{Data for the linear regressions, $\xfe = a + b \cdot \feh$,
	shown in Fig. \ref{fig:xfe-feh}.}
\label{table:fit.feh}
\setlength{\tabcolsep}{0.15cm}
\begin{tabular}{cccccc}
\hline\hline
\noalign{\smallskip}
	  El.   & $b$  & $\sigma_{\rm fit}$\,\tablefootmark{a} 
	& $\sigma_{\rm obs}$\,\tablefootmark{b}  
	& $\sigma_{\rm intrin}$\,\tablefootmark{c} 
	&  $N_{\star}$\,\tablefootmark{d}    \\ 
\noalign{\smallskip}
\hline
\noalign{\smallskip}
C  & $-0.034 \pm 0.056$ & 0.076   & 0.036 & $0.067 \pm 0.017$  & 22 \\
O  & $-0.069 \pm 0.042$ & 0.063   & 0.034 & $0.053 \pm 0.015$  & 23 \\
Mg & $-0.074 \pm 0.034$ & 0.050   & 0.031 & $0.039 \pm 0.013$  & 24 \\
Al & $+0.027 \pm 0.051$ & 0.070   & 0.044 & $0.054 \pm 0.018$  & 22 \\
Ca & $-0.119 \pm 0.026$ & 0.041   & 0.021 & $0.035 \pm 0.009$  & 25 \\
Sc & $+0.016 \pm 0.033$ & 0.048   & 0.024 & $0.043 \pm 0.011$  & 24 \\
Ti & $+0.001 \pm 0.026$ & 0.040   & 0.015 & $0.037 \pm 0.008$  & 25 \\
V  & $-0.023 \pm 0.034$ & 0.053   & 0.025 & $0.047 \pm 0.012$  & 25 \\
Cr & $-0.005 \pm 0.015$ & 0.022   & 0.019 & $0.011 \pm 0.005$  & 24 \\
Mn & $-0.014 \pm 0.031$ & 0.045   & 0.031 & $0.033 \pm 0.011$  & 24 \\
Co & $-0.210 \pm 0.017$ & 0.024   & 0.020 & $0.013 \pm 0.013$  & 24 \\
Ni & $-0.038 \pm 0.029$ & 0.036   & 0.026 & $0.025 \pm 0.009$  & 21 \\
Zn & $-0.187 \pm 0.032$ & 0.046   & 0.026 & $0.038 \pm 0.010$  & 24 \\
Y  & $+0.127 \pm 0.066$ & 0.093   & 0.022 & $0.090 \pm 0.020$  & 23 \\
Zr & $+0.064 \pm 0.072$ & 0.096   & 0.043 & $0.086 \pm 0.021$  & 22 \\
\noalign{\smallskip}
\hline
\end{tabular}
\tablefoot{\tablefoottext{a} {Standard deviation of the fit.}
\tablefoottext{b} {Observational error of \xfe .}
	\tablefoottext{c} {Intrinsic scatter derived from Eqs. (3) and (4).}
	\tablefoottext{d} {Number of stars in the fit.}}
\end{table}

\smallskip
The listed errors of the intrinsic dispersions in Table \ref{table:fit.feh}
were calculated as the quadratic sum of the statistical error of the standard deviation of
the fit ($\sigma_{\rm fit} / \sqrt{N_{\star}}$) and the error of
$\sigma_{\rm obs}$, which is dominated  by possible errors in
our estimate of the uncertainties of \teff , and \logg\ by say 30\%,
corresponding to errors of $\pm 40$\,K and $\pm 0.04$\,dex, respectively,
instead of the adopted errors of $\pm 30$\,K and $\pm 0.03$\,dex.
Adopting these errors of $\sigma_{\rm fit}$ and $\sigma_{\rm obs}$, there is a
3-$\sigma$ or more significant intrinsic dispersion for all elements except 
Cr, Co, and Ni as seen from Table \ref{table:fit.feh}. Notably, these elements may be
susceptible to 3D non-LTE effects (Sect. \ref{Other}), that are not accounted for here.

Our intrinsic dispersions are smaller than those obtained 
by \citet{griffith23} in their high-resolution abundance study of 86 subgiant stars with
metallicities in the range $-2.1 < \feh < -1.0$. Based on linear fits of \xfe\ as a function
of \feh , they find intrinsic dispersion for Mg, Ca, Sc, Ti, and V 
that are about a factor of two higher than
our values, see Table 4 of \citet{griffith23}, and for the iron-peak
elements the discrepancy is even higher. 

\begin{figure}
\centering
\includegraphics[width=8.5cm]{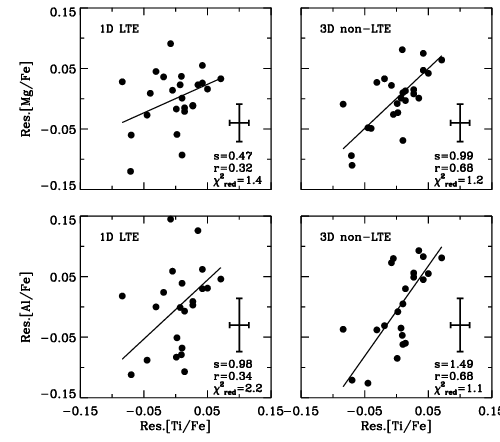}
\caption{Correlations of the residuals (individual star $-$ linear fit) 
of the \xfe -\feh\ relations for Mg, Al, and Ti
    using 1D LTE (left panels) and 3D non-LTE (right panels) abundances of Mg and Al.
The lines show maximum likelihood linear fits to the data taking into
account the errors in both coordinates.
The values of the slope, $s$, the Pearson correlation coefficient, $r$, and
the reduced chi-square of the fits are given.}
\label{fig:residuals-MgAl}
\end{figure}

During the analysis of the data shown in Fig. \ref{fig:xfe-feh}, we noted that the 
residuals (individual star $-$ linear fit) of the \xfe -\feh\ relations for 
Mg, Al, Ca, Sc, and Ti are correlated. As an example, Fig. \ref{fig:residuals-MgAl}
shows the residuals for Mg and Al versus the residuals for Ti.
After having applied 3N\,--\,1L corrections of the Mg and Al abundances,
one sees very clear correlations. We have, therefore, also determined dispersions 
relative to linear regressions
\begin{eqnarray}
\xfe = a + b \cdot \feh + c \cdot \alphafe \,,
\end{eqnarray}
where \alphafe\ is the mean value of \mgfe , \cafe , and \tife \,\footnote{When 
fitting \xfe\ for Mg, Ca, or Ti, the element itself
is not included in the calculation of \alphafe .}
The results are given in Table \ref{table:fit.alphafe}. The intrinsic dispersion is 
again determined from Eq. (3), where the error expected from observational errors is
now determined by the expression
\begin{eqnarray}
	\sigma_{\rm exp}^2 = \sigma_{\rm obs,[X/Fe]}^2 + b^2 \cdot \sigma_{\rm obs,[Fe/H]}^2
	+c^2 \cdot \sigma_{\rm obs,[\alpha/Fe]}^2,
\end{eqnarray}
where $\sigma_{\rm obs,[\alpha/Fe]} = 0.013$\,dex.

\smallskip
As seen from Table \ref{table:fit.alphafe}, the $c$ coefficients, that is the slopes
of \xfe\ versus \alphafe , have well defined values close to one for Mg, Al, Ca, Sc, and Ti,
and the intrinsic dispersions for these elements are much 
smaller than the corresponding
intrinsic dispersions for the $\xfe = a + b \cdot \feh$ fits given in Table \ref{table:fit.feh}.
This reflects the strong correlations between the residuals of the \xfe -\feh\ fits 
for the $\alpha$-elements as further discussed in Sect. \ref{SNIa}.
For Cr, Mn, Co, Ni, and Zn, the $c$ coefficients are not 
significantly different from zero indicating that there is no correlation between
their \xfe\ residuals and those of the $\alpha$-elements. These elements are, however,
highlighted in Sect. \ref{Other} as potentially having significant 3D non-LTE effects,
so the question of a correlation should be reconsidered, when 3N\,--\,1L corrections become available.

\begin{table}
\centering
	\caption[ ]{Data for the linear regressions, $\xfe = a + b \cdot \feh + c \cdot \alphafe$, 
	to our abundances.} 
\label{table:fit.alphafe}
\setlength{\tabcolsep}{0.15cm}
\begin{tabular}{cccccc}
\hline\hline
\noalign{\smallskip}
	El. & $c$  & $\sigma_{\rm fit}$\,\tablefootmark{a} 
	& $\sigma_{\rm obs}$\,\tablefootmark{b}  
	& $\sigma_{\rm intrin}$\,\tablefootmark{c} 
	&  $N_{\star}$\,\tablefootmark{d}    \\
\noalign{\smallskip}
\hline
\noalign{\smallskip}
C  & $+0.324 \pm 0.470$ & 0.078   & 0.036 & $0.069 \pm 0.018$  & 22 \\
O  & $+0.410 \pm 0.390$ & 0.065   & 0.034 & $0.055 \pm 0.016$  & 23 \\
Mg & $+0.923 \pm 0.195$ & 0.036   & 0.031 & $0.014 \pm 0.011$  & 24 \\
Al & $+1.273 \pm 0.324$ & 0.053   & 0.044 & $0.024 \pm 0.017$  & 22 \\
Ca & $+0.778 \pm 0.140$ & 0.027   & 0.021 & $0.010 \pm 0.008$  & 25 \\
Sc & $+0.937 \pm 0.172$ & 0.032   & 0.024 & $0.017 \pm 0.008$  & 24 \\
Ti & $+0.760 \pm 0.132$ & 0.026   & 0.015 & $0.015 \pm 0.006$  & 25 \\
V  & $+0.729 \pm 0.248$ & 0.046   & 0.025 & $0.037 \pm 0.010$  & 25 \\
Cr & $-0.062 \pm 0.120$ & 0.022   & 0.019 & $0.009 \pm 0.005$  & 24 \\
Mn & $-0.049 \pm 0.248$ & 0.046   & 0.031 & $0.034 \pm 0.011$  & 24 \\
Co & $+0.156 \pm 0.114$ & 0.021   & 0.020 & $0.006 \pm 0.013$  & 24 \\
Ni & $-0.197 \pm 0.218$ & 0.036   & 0.026 & $0.024 \pm 0.009$  & 21 \\
Zn & $-0.056 \pm 0.256$ & 0.047   & 0.026 & $0.039 \pm 0.010$  & 24 \\
Y  & $+1.334 \pm 0.485$ & 0.081   & 0.022 & $0.076 \pm 0.018$  & 23 \\
Zr & $+1.111 \pm 0.530$ & 0.088   & 0.043 & $0.077 \pm 0.020$  & 22 \\
\noalign{\smallskip}
\hline
\end{tabular}
\tablefoot{
\tablefoottext{a} {Standard deviation of the fit.}
\tablefoottext{b} {Observational error of \xfe .}
\tablefoottext{c} {Intrinsic scatter derived from Eqs. (3) and (6).}
\tablefoottext{d} {Number of stars in the fit.}}
\end{table}

\begin{figure}
\centering
\includegraphics[width=7.0cm]{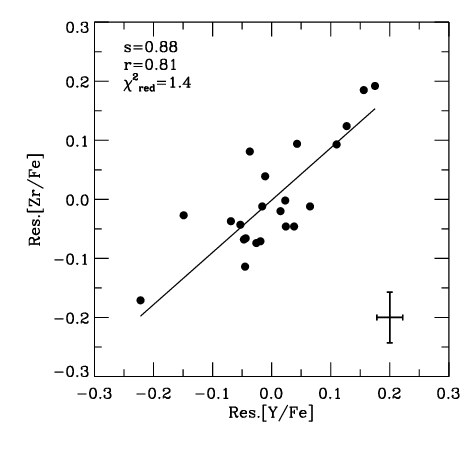}
	\caption{Correlation of the residuals of the  \xfe -\feh\ fits for Y and Zr.}
\label{fig:residuals-ZrY}
\end{figure}

As seen from Tables \ref{table:fit.feh} and \ref{table:fit.alphafe}, the intrinsic dispersion 
for the neutron capture elements, Y and Zr, is larger than for the other elements.
Interestingly, there is, however, a nice correlation
between the residuals of Y and Zr in the $\xfe = a + b \cdot \feh$ fits
as seen from Fig. \ref {fig:residuals-ZrY} and discussed in Sect. \ref{YZr}.

\section{Discussion}
\label{Discussion}

In this section we discuss possible explanations of the intrinsic dispersion
of abundance ratios at a given metallicity and the correlations of the residuals
in the \xfe\ - \feh\ relations.

\subsection{Binaries and peculiar stars}
\label{Binaries}
Three stars in our sample have been found to be single-lined spectroscopic
binaries, that is G\,126-62 and G\,59-27 \citep{latham02} and CD\,$-71 \, 1234$
\citep{ryan99}. None of them show particular large deviations
from the fitted \xfe\ - \feh\ relations, so the components are apparently too
faint to affect the derived abundance ratios significantly. These three stars have also
high values of the $Gaia$ RUWE parameter \citep{castro-ginard24}, i.e. RUWE $> 1.3$, indicating the presence
of unresolved components. Out of the remaining 22 stars in our sample only one, LP\,635-14,
has a RUWE parameter suggesting the existence of an unseen component, but again
there is no indication of effects on the derived \xfe\ ratios. Altogether, we conclude that
binarity of stars in our sample is unimportant, when discussing sources for the 
intrinsic dispersions in \xfe .

Stars with peculiar abundances of some elements are, however, important to identify.
In our sample, two stars,
G\,24-03 and HD\,160617, are know to belong to the class of nitrogen-rich 
dwarf stars.  \citet{spite22} determined abundances of six of such stars (including
G\,24-03 and HD\,160617) relative to six `normal' stars and found them to have $\nfe \simeq 1.0$
and to be somewhat underabundant in C and O and overabundant in Na and Al. Based
on this, they suggested that the N-rich dwarfs were originally born as second generation 
stars in globular clusters from gas enriched in N, Na, and Al through the
CNO, Ne-Na and Mg-Al cycles of hydrogen burning \citep{gratton19}.
Possible first-generation polluters of the gas are
intermediate-mass AGB stars \citep{ventura01} and fast-rotating massive stars
\citep{decressin07}. 
Our abundances of G\,24-03 and HD\,160617 agree with this scenario;
as seen from Fig. \ref{fig:xfe-feh} they stand out by having low values of 
\cfe\ and \ofe\ and high values of \alfe . Consequently, they were excluded when fitting
the \xfe\ - \feh\ trends for these elements. 

Another star having peculiar abundances is HD\,106038. It was shown by \citet{nissen97}
to have \xfe\ ratios for Si, Ni, Y, and Ba enhanced by more than 0.20\,dex
relative to `normal' halo stars with the same metallicity ($\feh \sim -1.4$).
Later \citet{smiljanic08} showed that the beryllium abundance of HD\,106038 is enhanced 
by a factor of 10 relative to halo dwarf stars with similar metalliicty and  evolutionary
stage and lithium is enhanced by a factor of two. They suggested
that these enhancements are due to spallation production of Li and Be in 
core-collapse SNe with exceptional high energy, so-called hypernovae \citep{fields02, nakamura04}.

As seen from Fig. \ref{fig:xfe-feh}, HD\,106038 stands out by being enhanced in \xfe\
for C, Mg, Al, Sc, Cr, Mn, Co, Ni, Zn, Y, and Zr by more than three times the standard
deviation of the linear fits to the \xfe\ - \feh\ trends. 
For these elements, HD\,106038 has been excluded from the fits.
According to the review by
\citet{nomoto13} only the enhancements of Co and Zn can be explained by
nucleosynthesis in hypernovae and a predicted enhancement of Ca and Ti is not seen in our abundances.
The origin of the abundance pattern of HD\,106038 is therefore unclear, but perhaps our precise 
abundances can help in finding a source. 

\subsection{Atomic diffusion effects}
Differences in atomic diffusion of elements may contribute to the dispersion 
of \xfe\ at a given metallicity even if the \teff -range of our stars is relatively small. 
As shown by \citet{dotter17}, models of old metal-poor
stars including atomic diffusion with radiative acceleration and turbulent mixing
predict that \ofe\ is decreased by $\sim 0.1$\,dex in turnoff stars relative to
\ofe\ in cooler main-sequence or subgiant stars, opposite to
\cafe\ that is enhanced in turnoff stars \citep[see Fig. 9, lower panel in][]{dotter17}.
Furthermore, \citet{korn07} and \citet{nordlander12}, in a study of elemental abundances
in turnoff and red giant stars belonging to the globular cluster NGC\,6397, found Fe to be 
more depleted than Ca and Ti in the turnoff stars, which they ascribe to
differential atomic diffusion between the elements. This interpretation was supported by
calculations based on the atomic diffusion models described in \citet{richard05}.

\begin{figure}
\centering
\includegraphics[width=8.0cm]{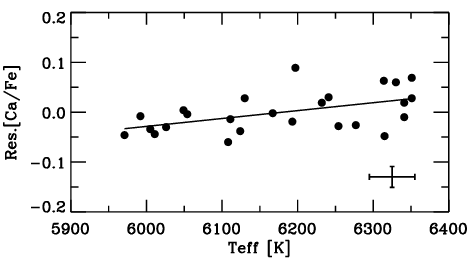}
\caption{Residuals in the \cafe -\feh\ fit versus \teff . 
The line shows the linear regression fit to the data.}
\label{fig:ResCaFe-Teff}
\end{figure}

In view of these results, we have investigated if there is a significant dependence of the 
residuals in the \xfe -\feh\ fits on effective temperature. As seen from Table \ref{table:fit.teff}
this is not the case for most elements, but for Ca, Sc, Ti, and V, the slope of
Res.\xfe\ versus \teff\ has a significance higher than 2-$\sigma$. As an example,
Fig. \ref{fig:ResCaFe-Teff} shows the trend of \cafe . From \teff \,=\,6000\,K to
\teff = 6300\,K, \cafe\ increases by 0.056\,dex. This is  qualitatively in
agreement with the predictions by \citet{dotter17}. Furthermore, we find a negative \ofe -\teff\
gradient, although only significant at the 1-$\sigma$ level, which also agrees with 
the Dotter et al. prediction.

\begin{table}
\centering
	\caption[ ]{Data for the linear regressions, Res.\,$\xfe = a + b \cdot \teff $,
	of the residuals in the \xfe -\feh\ fits.}
\label{table:fit.teff}
\setlength{\tabcolsep}{0.05cm}
\begin{tabular}{ccccc}
\hline\hline
\noalign{\smallskip}
	  El.   & $b$  & $\sigma_{\rm fit}$(Res.\xfe)\,\tablefootmark{a} 
	& $\sigma_{\rm fit}$(\xfe)\,\tablefootmark{b}  
	&  $N_{\star}$\,\tablefootmark{c}    \\ 
	& [dex/100 K]  &    &   &    \\
\noalign{\smallskip}
\hline
\noalign{\smallskip}
C  & $-0.0080 \pm 0.0138$ & 0.076   & 0.076 &  22 \\
O  & $-0.0094 \pm 0.0106$ & 0.062   & 0.063 &  23 \\
Mg & $+0.0065 \pm 0.0082$ & 0.049   & 0.050 &  24 \\
Al & $+0.0019 \pm 0.0127$ & 0.070   & 0.070 &  22 \\
Ca & $+0.0159 \pm 0.0056$ & 0.036   & 0.041 &  25 \\
Sc & $+0.0190 \pm 0.0069$ & 0.042   & 0.048 &  24 \\
Ti & $+0.0117 \pm 0.0058$ & 0.037   & 0.040 &  25 \\
V  & $+0.0181 \pm 0.0075$ & 0.047   & 0.053 &  25 \\
Cr & $+0.0012 \pm 0.0036$ & 0.022   & 0.022 &  24 \\
Mn & $+0.0097 \pm 0.0072$ & 0.043   & 0.045 &  24 \\
Co & $+0.0026 \pm 0.0040$ & 0.024   & 0.024 &  24 \\
Ni & $-0.0013 \pm 0.0065$ & 0.036   & 0.036 &  21 \\
Zn & $+0.0001 \pm 0.0077$ & 0.046   & 0.046 &  24 \\
Y  & $+0.0090 \pm 0.0156$ & 0.092   & 0.093 &  23 \\
Zr & $+0.0203 \pm 0.0158$ & 0.092   & 0.096 &  22 \\
\noalign{\smallskip}
\hline
\end{tabular}
\tablefoot{\tablefoottext{a} {Standard deviation for the Res.\,$\xfe = a + b \cdot \teff $ fit.}
\tablefoottext{b} {Standard deviation for the $\xfe = a + b \cdot \feh $ fit.}
\tablefoottext{c} {Number of stars in the fit.}}
\end{table}

The effect of differential atomic diffusion on the dispersion of \xfe\ versus \feh\
is, however, quite small. As seen from 
the two $\sigma_{\rm fit}$ columns in
Table \ref{table:fit.teff},
the standard deviations of the Res.\,$\xfe = a + b \cdot \teff $ fits for 
Ca, Sc, Ti, and V are about 10\% smaller than the corresponding standard deviations
of the $\xfe = a + b \cdot \feh $ fits and for the other elements there is no 
significant differences. Thus, we estimate that correction for atomic diffusion effects
decreases the derived intrinsic dispersions by less than about 10\%, which is smaller than 
the estimated error of $\sigma_{\rm intrin}$ in Table \ref{table:fit.feh}.

\subsection{Variations in the Ia\,/\,CC\,SN enrichment ratio}
\label{SNIa}
The strong correlation between the residuals of the \xfe -\feh\ fits for the 
$\alpha$-capture elements described in
Sect. \ref{results} suggests that the intrinsic dispersion of abundance ratios
at a given \feh\
is connected to variations in the ratio of element production by Type\,Ia and CC\,SNe
among star-forming regions.
This is the canonical explanation of the difference in \xfe\
between high-$\alpha$ and low-$\alpha$ stars found by NS10; the
high-$\alpha$ stars were made in-situ in our Galaxy with only CC\,SNe contributing to
the chemical evolution until a metallicity of $\feh \sim -0.7$ is reached, 
whereas the low-$\alpha$ stars were accreted from dwarf galaxies (primarily from the GSE galaxy) 
with a slower evolution allowing significant contributions from Type Ia\,SNe at $\feh \simgt -1.4$. 
We therefore see an increasing difference of \alphafe\ between high-$\alpha$ and low-$\alpha$ stars, 
when \feh\ increases from $-1.4$ to $-0.7$. There is, however, evidence that Type Ia\,SNe
start contributing elements at a lower metallicity than $\feh \simeq -1.4$ in 
surviving dwarf spheroidal galaxies such as Fornax \citep{hendricks14}, Sculptor \citep{hill19}, and 
Sextans \citep{theler20}. Furthermore, stars in some kinematic substructures of the
Galactic halo, whose progenitors are likely to be small dwarf galaxies, are found to 
have lower \alphafe\ than GSE stars in the metallicity range $-1.8 < \feh < -1.4$ by as much as 
0.15\,dex. This includes the prograde Helmi stream \citep{matsuno22a} and the
high energy, retrograde Sequoia structure \citep{matsuno22b, ceccarelli24}.

\begin{figure}
\centering
\includegraphics[width=8.5cm]{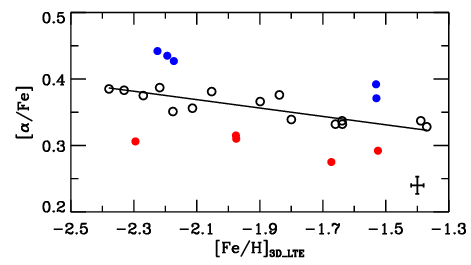}
\caption{\alphafe\ versus \feh . The line shows the linear regression to the whole sample.
Stars having positive residuals (stellar \alphafe\ $-$ linear fit)
three times larger than the estimated observational error of \alphafe\ ($\sigma_{\rm obs} = 0.013$\,dex)
are shown with filled blue circles 
and those with  $\Delta < 3 \sigma_{\rm obs}$ are shown with filled red circles
The rest of the stars are shown with open circles.}
\label{fig:alphafe}
\end{figure}

\begin{figure}
\centering
\includegraphics[width=8.5cm]{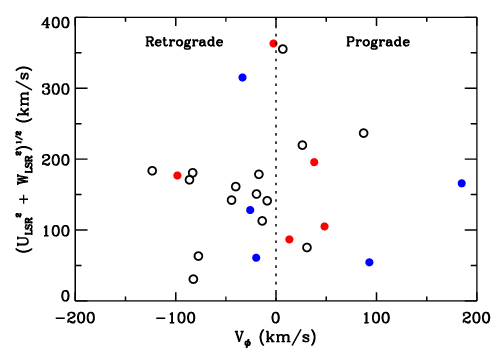}
\caption{Toomre diagram with the same symbols as in Fig. \ref{fig:alphafe}.}
\label{fig:Toomre}
\end{figure}

In view of these results, we have investigated if there is any correlation between 
\alphafe\ and the kinematics of our stars. Figure \ref{fig:alphafe} shows \alphafe\ as 
a function of \feh\ with stars deviating more than $\pm 3 \sigma_{\rm obs}$ from
the fitted linear relation shown with filled blue and red circles, respectively, whereas stars
in between are shown with open circles. Furthermore,  Fig. \ref{fig:Toomre} shows
the Toomre diagram with the
velocity components $U$, $V_{\phi}$, and $W$ calculated from $Gaia$
DR3 data as described in \citet{nissen21} and with the same symbols as in Fig. \ref{fig:alphafe}.
As seen, there is no obvious correlation between \alphafe\ enhancement or deficiency and 
the kinematics, but a larger sample of stars is needed to test for correlations.
 
\begin{figure}
\centering
\includegraphics[width=8.5cm]{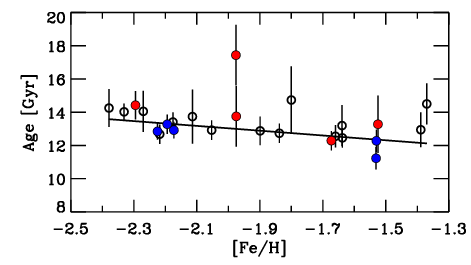}
\caption{Stellar age as a function of \feh\ with  the same symbols as in Fig. \ref{fig:alphafe}
and 1-sigma error bars on the ages. The line shows the linear regression fit to all stars
	using the inverse of the squared ager-error as weight.}
\label{fig:age-feh}
\end{figure}

We have also investigated if there is any  relation between \alphafe\ and stellar age
as determined by interpolating between 
Yonsei-Yale isochrones \citep{kim02} in the \teff - luminosity diagram onto 
the heavy element mass fraction $Z$ of the star calculated from the abundances
in Table \ref{table:A2}\,\footnote{For the elements N, Ne, and Si, which make a
significant contribution to $Z$ and are
missing in Table \ref{table:A2}, we assumed that \nefe\ and \sife\ are equal to \alphafe\ and
that \nfe = 0.0 except for the two nitrogen-rich stars,
G\,24-03   and HD\,160617, for which \nfe =1.0 was adopted.}.
The resulting ages are shown in Fig. \ref{fig:age-feh} as a function of \feh\ with
error bars estimated from the  1-$\sigma$ errors of \teff , luminosity, and $Z$. 
These error bars reflect the uncertainty 
on age differences relative to the standard star;
the absolute ages are more uncertain as they are more sensitive to parameters of the
stellar models such as the mixing length. We also note that the age error varies a lot;
it is large for stars on the main sequence, where the isochrones are tighter than near the turnoff.

The line shown in Fig. \ref{fig:age-feh} is a linear
regression fit to the ages using the inverse of the squared age error as weight.
The reduced chi-square of the fit is 0.92, so the age scatter can be explained by the estimated 
errors in the ages. Nevertheless, there is a tendency that 
stars having somewhat lower $\alpha$-element abundances than the majority of stars
(red circles)
fall above those stars having somewhat higher $\alpha$-element abundances
(blue circles). The same result is obtained when using PARSEC isochrones 
\citep{bressan12} to determine ages. The sign of this difference, which is significant
at the 2-sigma level, is surprising. 
If the stars with lower $\alpha$-element abundances were accreted from dwarf galaxies
with a low star-formation rate, one would expect them to be younger than 
the stars with higher $\alpha$-element abundances that are more likely to be
formed in-situ in regions with a high star-formation rate, provided that the star formation
started at the same time for the two populations. So, if the age difference between 
the stars of lower and higher $\alpha$-element abundances
shown in Fig. \ref{fig:age-feh} is confirmed for a larger sample, we must conclude 
that star formation in accreted dwarf galaxies started earlier than star formation in the
Galactic halo.

\begin{figure}
\centering
\includegraphics[width=8.5cm]{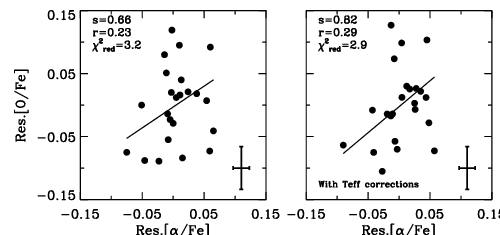}
\caption{Correlation of the residuals for the  \ofe -\feh\ and \alphafe -\feh\ fits.
In the right panel, corrections for the \teff\ dependence of the residuals have been applied.}
\label{fig:residuals-Oalpha}
\end{figure}

A problem of explaining the scatter of \alphafe\ at a given \feh\
as due to variations in the Ia\,/\,CC\,SNe enrichment ratio,
is a poor correlation between the residuals of \ofe\ and \alphafe .
Oxygen is thought to be made primarily in CC\,SNe like Mg, Ca, and Ti and a correlation
is therefore to be expected. As seen from  Fig. \ref{fig:residuals-Oalpha},
the correlation is, however, rather poor with a Pearson coefficient of only 0.23
and a reduced chi-square $\chi ^2_{\rm red} = 3.2$. If corrections for the \teff\
dependence of Res.\ofe\ and Res.\alphafe\ are included (see right panel of 
Fig. \ref{fig:residuals-Oalpha}), the correlation is somewhat improved, but is still
unconvincing. Hence, it seems that variations in the Ia\,/\,CC\,SNe enrichment ratio
cannot explain all of the intrinsic scatter of \ofe .

\subsection{Stochastic effects in sampling the IMF of CC\,SNe}
\label{Stochastic}
\begin{figure}
\centering
\includegraphics[width=8.0cm]{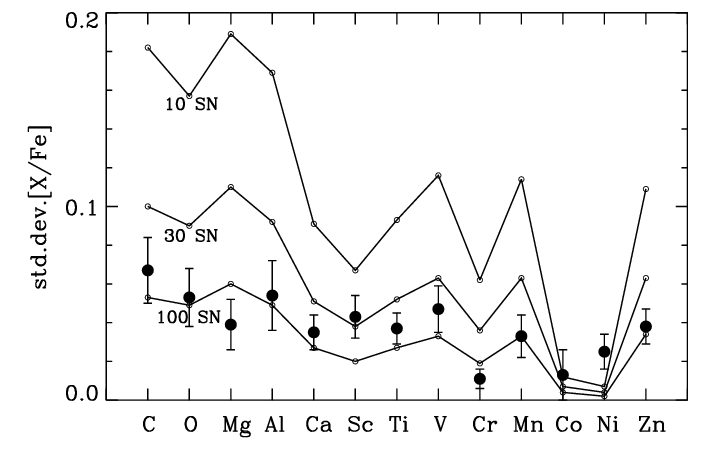}
\caption{Calculated standard deviation of \xfe\ as a function of the number of
CC\,SNe enriching a star-forming region (open circles connected by lines)
for `set M' yields from \citet{limongi18} corresponding to $\feh = -2.0$ and $\Vrot = 300$\,\kmprs\
in comparison with the intrinsic dispersion (filled circles with error bars)
of the \xfe -\feh\ trends.}
\label{fig:std.dev.}
\end{figure}

As the X/Fe yields of CC\,SNe have a strong dependence on stellar mass, stochastic
sampling of the IMF will introduce a scatter of 
abundance ratios at a given \feh\ between star-forming regions. Following the method described by 
\citet{griffith23}, we have calculated the standard deviation of \xfe\ as
a function of the number of CC\,SNe, $N_{\rm CCSN}$, enriching a star-forming region
by 10000 times repeated random sampling of  SNe with
progenitor masses between 13 and 80\,\Msun\ from a \citet{salpeter55} IMF weighted distribution;
see Fig. \ref{fig:std.dev.} as an example.
Yields corresponding to $\feh = -2.0$ were adopted from \citet{limongi18} using either their
`set M' yields, for which the whole mass of a SN is expelled, or their 'set R' yields, for 
which SNe with $M > 25 \Msun$ are assumed to collapse to a black hole and therefore
only contribute to the yields via stellar winds. In both cases, yields corresponding to initial 
stellar rotational velocities $\Vrot$ =  0, 150, and 300\,\kmprs\ are available.

As seen from Fig. \ref{fig:std.dev.}, the best match between predicted and measured
dispersions of \xfe\ is obtained for $N_{\rm CCSN}$ somewhere between 30 and 100.
Excluding Co and Ni, for which 
there is no significant variation in the calculated dispersion, and taking into account that the
calculated dispersion scales as $1 / \sqrt{N_{\rm CCSN}}$, the data in the figure 
suggest a weighted mean value $\langle N_{\rm CCSN} \rangle  = 64 \pm 14$. A similar number 
is obtained for `set M' yields corresponding
to $\Vrot$ =  0 and 150\,\kmprs  and also if yields corresponding to $\feh = -3.0$ are used.
'Set R' yields suggest a somewhat smaller number, typically 
$N_{\rm CCSN} \simeq 50$. A similar number was obtained by \citet{griffith23}. Although
their derived intrinsic dispersions are about a factor of two higher than ours, 
this is compensated by their use of solar metallicity yields from \citet{sukhbold16},
which lead to larger standard deviations of \xfe\ for a given number of SNe
than the yields of \citet{limongi18}.
As discussed by \citet[][see their Eq.(6)]{griffith23}, the  region over which the products of 
50 CC\,SNe are mixed has to have a gas mass of about $10^5$\,\Msun\ in order  
to reach an oxygen abundance similar to our stars. Interestingly, this is a typical
mass of giant molecular clouds \citep{fukui10}.

\subsection{Yttrium and Zirconium}
\label{YZr}
Y and Zr belong to the group of light neutron-capture elements, which also includes strontium.
At the metallicities of our stars, these elements are probably made by both the
$r$-process and the weak $s$-process in massive stars, but some contribution from the
main $s$-process in low-mass ($M = 1-3 \Msun$) AGB stars cannot be excluded
\citep{cristallo11}.
As shown in Table \ref{table:fit.feh}, the intrinsic dispersion for \yfe\ and \zrfe\ as a function of
\feh\ is about a factor of two higher than the dispersion for the $\alpha$-elements.
The same result has been obtained for a larger sample of halo stars stars with $-3.5 < \feh < -2.0$ 
by \citet{li22}. They also find a strong correlation between the residuals for Y and Zr
like in our Fig. \ref{fig:residuals-ZrY}, indicating that the two elements were made 
by the same nucleosynthetic processes. It should also be mentioned that \citet{griffith26} 
recently measured the
dispersion of several neutron-capture elements in their sample of 86 subgiant stars
with $-2.0 < \feh < -1.0$. For \yfe\ and \zrfe , they find intrinsic dispersions of 
0.20\,dex and 0.17\,dex, respectively, which is about a factor of two higher than the value of 0.09\,dex
we are finding.

Based on the stochastic chemical evolution model of \citet{cescutti08},
\citet{cescutti14} have shown that the large scatter of neutron-capture elements
in metal-poor halo stars can be explained by combining the weak $s$-process in fast rotating 
stars with various scenarios for the $r$-process in massive stars. 
\citet{scannapieco22} have simulated these processes in a hydro-dynamical cosmological
model, where the formation of the $\alpha$ elements is also included, and where
accretion from dwarf galaxies is taken into account. Interestingly, the model
predicts that the ratio of the scatter of \srfe\ is about twice as large as the
scatter of \mgfe\ \citep[see Fig. 9 in][]{scannapieco22} in agreement with the ratios of
the intrinsic dispersions in our Table \ref{table:fit.feh}.

\section{Summary and conclusions}
\label{Conclusions}

Based on VLT/UVES blue spectra, we have determined abundance ratios in
a sample of 25 halo stars with $-2.4 < \feh < -1.3$ belonging to the turnoff region of the HR-diagram.
As the spectra have S/N$\simgt 200$ and 
\teff\ was obtained from profiles of the H$\beta$ line and \logg\ via $Gaia$ parallaxes,
the error of differential values of \xfe\ could be held at a level of 0.02\,-\,0.04\,dex,
which allowed us to determine reliable values of the intrinsic dispersion in 
linear fits of \xfe\ as a function of \feh . The size of the derived dispersions are smaller
than found in previous investigations \citep{belokurov22, griffith23, griffith26}
and correlations between the residuals in 
the fits for different elements provide interesting constraints on possible explanations 
of the intrinsic dispersions.

Abundances were derived through a 1D LTE model atmosphere analysis of
measured  equivalent widths of spectral lines. 3D non-LTE corrections were applied for
C, O, Mg, Al, and Ca for which only lines from neutral atoms were available and 
3D LTE corrections were applied for Ti and Fe, 
for which lines from ionized atoms could be used. The corrections have important effects on the slopes
of \xfe -\feh\ relations but only small effects on the scatter at a given \feh , which
can be ascribed to the relatively small range in \teff\ and \logg\ for our sample of stars.
For Mg and Al, the 3D non-LTE corrections are, however, important for the correlation
of the residuals in the \xfe -\feh\ fits with residuals for other elements
(Fig. \ref{fig:residuals-MgAl}). Hence, it would be important
to obtain 3D non-LTE corrections for the remaining elements, especially Mn, Co, Ni, and Zn,
for which only lines from neutral atoms are available. 

Correlations of the residuals in the \xfe -\feh\ fits with effective temperature
suggest that differential  atomic diffusion including radiative acceleration has a significant
effect on \cafe , \scfe , \tife , and \vfe . The effect is, however, small for the narrow
\teff\ range of our stars and gives at most a 10\% contribution to the intrinsic scatter
of the abundance ratios.

Variations in the Ia\,/\,CC\,SNe enrichment ratio between star-forming regions in the
Milky Way halo and dwarf galaxies is a possible reason for  the intrinsic scatter
in the \xfe -\feh\ relations.
This is indeed the canonical explanation of the 
striking difference between high-$\alpha$ and
low-$\alpha$ stars for metallicities $\feh > -1.3$.
At lower metallicities, such as considered here,
there is evidence that
variations in \alphafe\ are still significant;
this may extend down to $\feh \sim -2.0$ for stars
with extreme kinematics \citep{matsuno22a, matsuno22b,ceccarelli24}. 
Correlations between the residuals of \xfe\
for Mg, Al, Ca, and Ti (Table \ref{table:fit.alphafe} and Fig. \ref{fig:residuals-MgAl}) 
support this scenario, but the lack of a clear correlation between 
the residuals for oxygen and the $\alpha$-elements (Fig. \ref{fig:residuals-Oalpha}) 
is puzzling and there is no obvious
correlation between the kinematics of our stars and \alphafe\ (Fig. \ref{fig:Toomre}). 
Clearly, a larger sample of halo stars with high-precision abundances is
needed to study these problems.

Another possibility for explaining the intrinsic scatter in the \xfe -\feh\ relations
is stochastic effects in sampling the IMF of CC\,SNe in star-forming regions. 
Such effects were also discussed by \citet{nissen94}, who used the upper
limits for the scatter of \ofe , \mgfe , and \cafe\ in nine metal-poor  halo stars
to estimate that $N_{\rm CCSN} \simgt 20$ is
required to get scatters in agreement with those observed. In Sect. \ref{Stochastic},
yields from \citet{limongi18} were used to estimate that $N_{\rm CCSN} \sim$\,50-70
is required to explain the small intrinsic scatter of \xfe\ we have determined 
(Fig. \ref{fig:std.dev.}). A similar number of 
CC\,SNe was estimated \citet{griffith23} from the scatter of abundance ratios
in 86 subgiant stars, and as shown by them this implies
that the CC\,SNe ejecta are mixed over a gas mass of $\sim \! 10^5 \Msun$, which is the
typical mass of giant molecular clouds. 

Among the iron-peak elements, Cr, Mn, Ni, Co, and Zn,
\citet{nissen10, nissen11} found differences in \nife\ and \znfe\ between high-$\alpha$ 
and low-$\alpha$ halo stars at $\feh > -1.3$. For the present sample of
more metal-poor stars, we find, however, no correlation of \nife\ and \znfe\
with \alphafe\ (Table \ref{table:fit.alphafe} in comparison with Table \ref{table:fit.feh}). 
On the other hand, we note that predictions of the stochastic effect for iron-peak elements 
agree with the derived intrinsic dispersion of the \xfe -\feh\ fits
in the sense that higher values are predicted for Mn and Zn than for Cr,  Co, and Ni
(Fig. \ref{fig:std.dev.}). 3D non-LTE corrections for the iron-peak
elements should, however, be taken into account, before conclusions can be drawn, 

It is likely that both variations in the Ia\,/\,CC\,SNe enrichment ratio and 
stochastic effects in sampling the IMF contribute to the scatter in the \xfe -\feh\ relations.
The small intrinsic scatter, we have determined for the metallicity range $-2.4 < \feh < -1.3$
sets in any case important constraints on these effects, and it also constrains other possible 
sources of the scatter, such as differences in the IMF between star forming regions
\citep{bastian10} and bursts of star formation \citep{gilmore91}.

\begin{acknowledgements}
We thank Bengt Gustafsson for very useful comments on a first version of this paper,
and the referee is thanked for many detailed suggestions that helped to clarify the text. 
AMA acknowledges support from the Swedish Research Council (VR 2020-03940, VR 2025-05167), 
and the Crafoord Foundation via the Royal Swedish Academy of Sciences (CR 2024-0015).  	
This research was supported by computational resources provided by 
the Australian Government through the National Computational Infrastructure (NCI)
under the National Computational Merit Allocation Scheme and
the ANU Merit Allocation Scheme (project y89).  The computations were also made
possible thanks to resources at the National Supercomputing Centre
(NSC, Tetralith cluster) provided by the National Academic Infrastructure
for Supercomputing in Sweden (NAISS), partially funded by the
Swedish Research Council through grant agreement no. 2022-06725.
This research has made use of data from the European Space Agency (ESA) mission
{\it Gaia} (\url{https://www.cosmos.esa.int/gaia}), processed by the {\it Gaia}
Data Processing and Analysis Consortium (DPAC,
\url{https://www.cosmos.esa.int/web/gaia/dpac/consortium}). Funding for the DPAC
has been provided by national institutions, in particular the institutions
participating in the {\it Gaia} Multilateral Agreement.
This research has made use of the SIMBAD database operated at CDS, Strasbourg, 
France \citep{wenger00}.
\end{acknowledgements}

\bibliographystyle{aa}
\bibliography{nissen.2026}

\begin{appendix} 

\nopagebreak

\begin{table*}
\section{Atomic data and derived abundances}

\centering
	\caption[ ]{List of spectral lines.} 
\label{table:linelist}
\setlength{\tabcolsep}{0.15cm}
\begin{tabular}{rccrrcrcrccrrcr}
\noalign{\smallskip}
\hline\hline
\noalign{\smallskip}
Elem. & $\lambda$&Exc.pot.&log$gf$  & EW\,\tablefootmark{a}    & W\,\tablefootmark{b}& Ref. & \hspace{1cm}  &
Elem. & $\lambda$&Exc.pot.&log$gf$  & EW\,\tablefootmark{a}    & W\,\tablefootmark{b}& Ref. \\
& [\AA ] &  [eV]  &         &  [m\AA ]&                     &      &            &
& [\AA ] &  [eV]  &         &  [m\AA ]&                     &       \\     
\noalign{\smallskip}
\hline
\noalign{\smallskip}
  \MgI &  4571.10 &   0.00 & $-5.732$ &    21.7 & U & 1 &    & \FeII &  4178.86 &   2.58 & $-2.510$ &    42.7& A & 13\\
    -  &  4703.00 &   4.34 & $-0.456$ &   104.1 & A & 2 &    &    -  &  4233.17 &   2.58 & $-1.970$ &    70.2& A & 13\\
  \AlI &  3961.53 &   0.01 & $-0.333$ &   115.8 & A & 3 &    &    -  &  4416.83 &   2.78 & $-2.650$ &    32.0& A & 13\\
  \CaI &  4425.44 &   1.88 & $-0.393$ &    59.6 & A & 4 &    &    -  &  4489.18 &   2.83 & $-2.960$ &    19.7& A & 13\\
    -  &  4435.69 &   1.89 & $-0.528$ &    52.9 & A & 4 &    &    -  &  4491.41 &   2.85 & $-2.710$ &    23.6& A & 13\\
    -  &  4526.93 &   2.71 & $-0.548$ &    16.0 & U & 4 &    &    -  &  4508.29 &   2.85 & $-2.440$ &    39.4& A & 13\\
    -  &  4578.56 &   2.52 & $-0.697$ &    16.7 & U & 4 &    &    -  &  4515.34 &   2.84 & $-2.600$ &    33.6& A & 13\\
    -  &  4585.87 &   2.52 & $-0.338$ &    32.9 & U & 4 &    &    -  &  4520.23 &   2.81 & $-2.650$ &    31.1& A & 13\\
 \ScII &  4400.40 &   0.61 & $-0.540$ &    45.4 & U & 5 &    &    -  &  4541.52 &   2.85 & $-2.980$ &    16.2& A & 13\\
    -  &  4415.56 &   0.60 & $-0.680$ &    40.1 & U & 5 &    &    -  &  4555.89 &   2.83 & $-2.400$ &    40.7& A & 13\\
    -  &  4670.41 &   1.36 & $-0.600$ &    13.3 & U & 5 &    &    -  &  4576.34 &   2.84 & $-2.950$ &    15.6& A & 13\\
 \TiII &  4028.35 &   1.89 & $-0.920$ &    43.0 & U & 6 &    &    -  &  4582.83 &   2.84 & $-3.180$ &    10.8& A & 13\\
    -  &  4184.31 &   1.08 & $-2.490$ &    15.3 & U & 6 &    &    -  &  4583.84 &   2.81 & $-1.930$ &    64.8& A & 13\\
    -  &  4394.07 &   1.22 & $-1.770$ &    37.7 & U & 6 &    &    -  &  4620.52 &   2.83 & $-3.210$ &     8.7& A & 13\\
    -  &  4395.85 &   1.24 & $-1.930$ &    27.3 & U & 6 &    &  \CoI &  4121.33 &   0.92 & $-0.330$ &    44.6& A & 14\\
    -  &  4399.78 &   1.24 & $-1.200$ &    64.7 & U & 6 &    &  \NiI &  4470.48 &   3.40 & $-0.300$ &    10.3& A & 15\\
    -  &  4417.72 &   1.16 & $-1.430$ &    66.9 & U & 6 &    &    -  &  4605.00 &   3.48 & $-0.240$ &    10.0& A & 15\\
    -  &  4418.34 &   1.24 & $-1.990$ &    27.0 & U & 6 &    &    -  &  4648.66 &   3.42 & $-0.090$ &    14.0& A & 15\\
    -  &  4444.56 &   1.12 & $-2.200$ &    22.3 & U & 6 &    &    -  &  4715.77 &   3.54 & $-0.330$ &     7.4& A & 15\\
    -  &  4450.49 &   1.08 & $-1.520$ &    56.6 & U & 6 &    &    -  &  4756.52 &   3.48 & $-0.270$ &     9.0& A & 15\\
    -  &  4464.46 &   1.16 & $-2.080$ &    38.6 & U & 7 &    &    -  &  4786.54 &   3.42 & $-0.180$ &    13.6& A & 15\\
    -  &  4470.86 &   1.16 & $-2.280$ &    22.5 & U & 7 &    &    -  &  4829.03 &   3.54 & $-0.330$ &     9.3& A & 16\\
    -  &  4488.33 &   3.12 & $-0.500$ &    14.2 & U & 6 &    &    -  &  4831.18 &   3.61 & $-0.320$ &     7.2& A & 15\\
    -  &  4544.02 &   1.24 & $-2.410$ &     8.2 & U & 8 &    &    -  &  4904.42 &   3.54 & $-0.170$ &    12.1& A & 16\\
    -  &  4657.20 &   1.24 & $-2.150$ &    15.5 & U & 8 &    &    -  &  4918.37 &   3.84 & $-0.230$ &     6.1& A & 16\\
    -  &  4708.67 &   1.24 & $-2.350$ &    14.2 & U & 6 &    &  \ZnI &  4722.16 &   4.03 & $-0.390$ &    11.9& A & 17\\
    -  &  4798.54 &   1.08 & $-2.660$ &    10.6 & U & 6 &    &    -  &  4810.54 &   4.08 & $-0.170$ &    15.8& A & 17\\
  \VII &  3951.96 &   1.48 & $-0.730$ &    19.9 & U & 9 &    &  \YII &  3788.70 &   0.10 & $-0.066$ &    42.7& U & 18\\
    -  &  4005.71 &   1.82 & $-0.450$ &    22.4 & U & 9 &    &    -  &  3950.36 &   0.10 & $-0.488$ &    27.1& U & 18\\
    -  &  4023.38 &   1.80 & $-0.610$ &    16.5 & U & 9 &    &    -  &  4398.02 &   0.13 & $-0.999$ &    12.2& U & 18\\
 \CrII &  4558.65 &   4.07 & $-0.430$ &    29.0& U & 10 &    &    -  &  4883.69 &   1.08 & $ 0.070$ &    16.4& U & 18\\
    -  &  4588.20 &   4.07 & $-0.650$ &    21.5& U & 10 &    &    -  &  4900.12 &   1.03 & $-0.090$ &    13.2& U & 18\\
    -  &  4634.08 &   4.07 & $-1.050$ &    11.3& U & 10 &    & \ZrII &  4208.98 &   0.71 & $-0.510$ &    14.8& U & 19\\
    -  &  4824.14 &   3.87 & $-0.920$ &    20.4& U & 10 &    &       &          &        &          &            &   \\
  \MnI &  4055.55 &   2.14 & $-0.070$ &    15.3& A & 11 &    &       &          &        &          &            &   \\
    -  &  4082.94 &   2.18 & $-0.354$ &     8.1& A & 11 &    &       &          &        &          &            &   \\
    -  &  4754.04 &   2.28 & $-0.085$ &    12.9& A & 11 &    &       &          &        &          &            &   \\
    -  &  4762.38 &   2.89 & $ 0.426$ &    10.9& A & 11 &    &       &          &        &          &            &   \\
    -  &  4766.42 &   2.92 & $ 0.100$ &     5.9& A & 12 &    &       &          &        &          &            &   \\
    -  &  4783.42 &   2.30 & $ 0.042$ &    15.7& A & 11 &    &       &          &        &          &            &   \\
    -  &  4823.51 &   2.32 & $ 0.144$ &    17.9& A & 11 &    &       &          &        &          &            &   \\
\noalign{\smallskip}
\hline
\end{tabular}
	\tablefoot{\tablefoottext{a}{Equivalent width for the standard star HD\,110621.}
	\tablefoottext{b} {Van der Waals broadening: (A) Anstee, Barklem, O'Mara calculations
\citep{barklem16}; (U) Uns{\"o}ld approximation with enhancement factor 1.5 \citep{unsold55}.}}

\tablebib{
	(1)~\citet{jonsson97};
	(2)~\citet{pehlivan17};
	(3)~\citet{nist24};
	(4)~\citet{denhartog21};
	(5)~\citet{lawler19};
	(6)~\citet{wood13};
	(7)~\citet{roberts73};
	(8)~\citet{kostyk83};
	(9)~\citet{wood14a};
	(10)~\citet{lawler17};
	(11)~\citet{booth84};
	(12)~\citet{greenlee79};
	(13)~\citet{melendez09};
	(14)~\citet{lawler15};
	(15)~\citet{wood14b};
	(16)~\citet{kostyk82};
	(17)~\citet{biemont80};
	(18)~\citet{hannaford82};
	(19)~\citet{ljung06}.
}

\end{table*}

\begin{sidewaystable*}
	\centering
	\caption[ ]{Atmospheric parameters and abundance ratios with 1D/3D (non-)LTE corrections
	applied as indicated in Fig. \ref{fig:xfe-feh}.}
\label{table:A2}
\setlength{\tabcolsep}{0.10cm}
\begin{tabular}{rccccrrrrrrrrrrrrrrr}
\noalign{\smallskip}
\hline\hline
\noalign{\smallskip}
	Star & \teff & \logg & \FeH & \turb & \CFe & \OFe & \MgFe & \AlFe & \CaFe & \ScFe & \TiFe & \VFe & \CrFe & \MnFe & \CoFe & \NiFe & \ZnFe & \YFe & \ZrFe  \\
	& K &  &  & \kmprs & &  & &  &  &  &  &  &  &  &  &  &  &  & \\
\noalign{\smallskip}
\hline
\noalign{\smallskip}
CD$-$30\,18140 & 6241 &  4.15 &$-1.838$ &  1.40 &$ 0.008$ & 0.634 & 0.368 &$-0.306$ & 0.420 &$ 0.103$ & 0.342 & 0.180 &$-0.007$ &$-0.492$ &$ 0.004$ &$-0.180$ &$-0.009$ &$-0.035$ & 0.349\\
CD$-$35\,14849 & 6254 &  4.30 &$-2.270$ &  1.09 &$ 0.079$ & 0.620 & 0.418 &$-0.405$ & 0.414 &$ 0.047$ & 0.296 & 0.185 &$ 0.016$ &$-0.514$ &$ 0.123$ &$-0.057$ &$ 0.076$ &$-0.281$ & 0.200\\
CD$-$42\,14278 & 6011 &  4.38 &$-1.976$ &  1.15 &$ 0.046$ & 0.535 & 0.354 &$-0.403$ & 0.363 &$-0.026$ & 0.231 & 0.102 &$ 0.003$ &$-0.497$ &$ 0.000$ &$-0.128$ &$ 0.052$ &$-0.071$ & 0.200\\
 CD$-$71\,1234 & 6315 &  4.24 &$-2.295$ &  1.37 &$-0.026$ & 0.570 & 0.277 &$-0.496$ & 0.397 &$ 0.024$ & 0.245 & 0.199 &$-0.045$ &$-0.546$ &$ 0.085$ &         &$ 0.142$ &         &      \\
 CS\,22943-095 & 6314 &  4.16 &$-2.194$ &  1.38 &$ 0.142$ & 0.730 & 0.454 &$-0.289$ & 0.496 &$ 0.079$ & 0.357 & 0.248 &$-0.036$ &$-0.540$ &$ 0.087$ &$-0.128$ &$ 0.150$ &$ 0.033$ & 0.417\\
      G\,04-37 & 6277 &  4.27 &$-2.379$ &  1.30 &$ 0.031$ & 0.770 & 0.403 &$-0.372$ & 0.429 &$ 0.032$ & 0.325 & 0.234 &$-0.013$ &$-0.458$ &$ 0.145$ &         &$ 0.168$ &$-0.184$ &      \\
      G\,11-44 & 6108 &  4.39 &$-1.975$ &  1.15 &$ 0.055$ & 0.623 & 0.315 &$-0.492$ & 0.347 &$ 0.030$ & 0.270 & 0.219 &$ 0.007$ &$-0.398$ &$ 0.085$ &$-0.061$ &$ 0.174$ &$-0.317$ & 0.075\\
      G\,13-09 & 6330 &  4.02 &$-2.225$ &  1.40 &$ 0.100$ & 0.599 & 0.446 &$-0.292$ & 0.496 &$ 0.154$ & 0.386 & 0.322 &$-0.035$ &$-0.494$ &$ 0.074$ &$-0.157$ &$ 0.086$ &$-0.089$ & 0.184\\
      G\,18-39 & 6049 &  4.24 &$-1.389$ &  1.20 &$ 0.165$ & 0.623 & 0.321 &$-0.257$ & 0.341 &$ 0.099$ & 0.351 & 0.211 &$-0.010$ &$-0.528$ &$-0.065$ &$-0.126$ &$ 0.055$ &$ 0.106$ & 0.408\\
      G\,20-08 & 6124 &  4.38 &$-2.114$ &  1.20 &$ 0.170$ & 0.712 & 0.400 &$-0.408$ & 0.385 &$-0.019$ & 0.284 & 0.074 &$-0.037$ &$-0.501$ &$ 0.041$ &$-0.120$ &$ 0.128$ &$-0.158$ & 0.123\\
      G\,24-03 & 6026 &  4.37 &$-1.525$ &  1.09 &$-0.261$ & 0.373 & 0.281 &$ 0.023$ & 0.323 &$ 0.039$ & 0.275 & 0.175 &$ 0.003$ &$-0.500$ &$-0.081$ &$-0.142$ &$ 0.013$ &$-0.085$ & 0.207\\
      G\,29-23 & 6167 &  4.07 &$-1.638$ &  1.37 &$-0.086$ & 0.545 & 0.315 &$-0.365$ & 0.365 &$ 0.096$ & 0.317 & 0.220 &$-0.050$ &$-0.522$ &$-0.068$ &$-0.179$ &$-0.065$ &$-0.089$ & 0.349\\
      G\,59-27 & 6232 &  4.22 &$-1.899$ &  1.41 &$ 0.144$ & 0.713 & 0.355 &$-0.424$ & 0.417 &$ 0.027$ & 0.329 & 0.195 &$-0.029$ &$-0.496$ &$-0.008$ &$-0.140$ &$ 0.051$ &$-0.096$ & 0.290\\
     G\,126-52 & 6341 &  4.16 &$-2.176$ &  1.30 &$-0.013$ & 0.548 & 0.309 &$-0.433$ & 0.421 &$ 0.075$ & 0.325 & 0.217 &$-0.008$ &$-0.400$ &$ 0.052$ &$-0.079$ &$ 0.101$ &$-0.174$ & 0.190\\
     G\,126-62 & 6197 &  4.08 &$-1.531$ &  1.37 &$-0.080$ & 0.519 & 0.411 &$-0.401$ & 0.443 &$ 0.088$ & 0.324 & 0.212 &$-0.018$ &$-0.435$ &$-0.036$ &$-0.130$ &$-0.027$ &$-0.058$ & 0.203\\
     HD\,84937 & 6341 &  4.13 &$-2.053$ &  1.33 &$ 0.011$ & 0.544 & 0.382 &$-0.338$ & 0.435 &$ 0.109$ & 0.329 & 0.202 &$-0.015$ &$-0.480$ &$ 0.026$ &$-0.195$ &$ 0.076$ &$-0.131$ & 0.167\\
    HD\,106038 & 5992 &  4.26 &$-1.370$ &  1.18 &$ 0.258$ & 0.593 & 0.469 &$ 0.396$ & 0.327 &$ 0.200$ & 0.328 & 0.257 &$ 0.068$ &$-0.379$ &$ 0.030$ &$ 0.131$ &$ 0.124$ &$ 0.435$ & 0.647\\
    HD\,108177 & 6111 &  4.29 &$-1.639$ &  1.18 &$ 0.029$ & 0.620 & 0.339 &$-0.392$ & 0.353 &$ 0.072$ & 0.322 & 0.174 &$-0.021$ &$-0.533$ &$-0.023$ &$-0.128$ &$ 0.027$ &$-0.068$ & 0.255\\
    HD\,110621 & 6130 &  4.12 &$-1.530$ &  1.30 &$ 0.040$ & 0.610 & 0.377 &$-0.309$ & 0.382 &$ 0.130$ & 0.357 & 0.220 &$-0.019$ &$-0.509$ &$-0.056$ &$-0.127$ &$-0.008$ &$ 0.071$ & 0.368\\
    HD\,160617 & 6005 &  3.91 &$-1.673$ &  1.39 &$-0.118$ & 0.336 & 0.247 &$-0.012$ & 0.337 &$-0.017$ & 0.244 & 0.124 &$-0.036$ &$-0.507$ &$-0.060$ &$-0.155$ &$-0.065$ &$-0.101$ & 0.199\\
    HD\,181743 & 5971 &  4.41 &$-1.800$ &  1.13 &$ 0.094$ & 0.662 & 0.372 &$-0.288$ & 0.340 &$ 0.031$ & 0.307 & 0.131 &$-0.026$ &$-0.562$ &$-0.007$ &$-0.149$ &$ 0.027$ &$-0.058$ & 0.237\\
    HD\,188031 & 6193 &  4.23 &$-1.660$ &  1.26 &$-0.030$ & 0.587 & 0.332 &$-0.442$ & 0.350 &$ 0.076$ & 0.316 & 0.186 &$-0.007$ &$-0.506$ &$-0.040$ &$-0.153$ &$ 0.002$ &$-0.124$ & 0.229\\
    HD\,215801 & 6054 &  3.84 &$-2.218$ &  1.30 &$-0.057$ & 0.656 & 0.389 &$-0.324$ & 0.432 &$ 0.087$ & 0.342 & 0.181 &$-0.047$ &$-0.581$ &$ 0.057$ &$-0.163$ &$ 0.074$ &$-0.061$ & 0.218\\
    HD\,338529 & 6351 &  4.11 &$-2.173$ &  1.36 &$ 0.048$ & 0.644 & 0.420 &$-0.316$ & 0.499 &$ 0.123$ & 0.365 & 0.221 &$-0.020$ &$-0.495$ &$ 0.101$ &$-0.090$ &$ 0.101$ &$-0.097$ & 0.231\\
    LP\,635-14 & 6351 &  4.15 &$-2.331$ &  1.50 &$ 0.101$ & 0.618 & 0.364 &$-0.296$ & 0.477 &$ 0.059$ & 0.310 & 0.262 &$ 0.040$ &$-0.467$ &$ 0.109$ &         &$ 0.191$ &$ 0.035$ & 0.415\\
\noalign{\smallskip}
\hline
\end{tabular}

\end{sidewaystable*}

\end{appendix}

\end{document}